\documentstyle[epsfig]{mn}
\input{epsf}

\renewcommand{\d}{{\rm d}}
\newcommand{\beq}{\begin{equation}}
\newcommand{\eeq}{\end{equation}}
\newcommand{\beqa}{\begin{eqnarray}} 
\newcommand{\eeqa}{\end{eqnarray}}
\newcommand{\bea}{\begin{array}} 
\newcommand{\ea}{\end{array}} 
\newcommand{\lag}{\langle}
\newcommand{\rag}{\rangle}
\newcommand{\Om}{\Omega_{\rm m}}
\newcommand{\Ol}{\Omega_{\Lambda}}
\newcommand{\De}{{\cal D}}
\newcommand{\gam}{\gamma}
\newcommand{\cP}{{\cal P}}
\newcommand{\kappamin}{\kappa_{\rm min}}
\newcommand{\kappah}{\hat{\kappa}}
\newcommand{\wh}{\hat{w}}
\newcommand{\kaphs}{\hat{\kappa}_s}
\newcommand{\Map}{M_{\rm ap}}
\newcommand{\tS}{\tilde{S}}
\newcommand{\bx}{{\bf x}}
\newcommand{\bk}{{\bf k}}
\newcommand{\kpar}{k_{\parallel}}
\newcommand{\kperp}{\bk_{\perp}}
\newcommand{\kperpDt}{k_{\perp}\De\theta_s}

\newcommand{\gamhs}{\hat{\gamma}_s}
\newcommand{\inta}{\int_{-i\infty}^{+i\infty}}

\newcommand{\xib}{\overline{\xi}}
\newcommand{\phikap}{\varphi_{\kaphs}}
\newcommand{\gamonehs}{\hat{\gamma}_{1s}}

\newcommand{\Igamone}{I_{\gam_1}}
\newcommand{\phigamone}{\varphi_{\gamonehs}}

\newcommand{\ysgamone}{y_{s,\gamonehs}}

\newcommand{\dum}{s}
\newcommand{\xidum}{\xi_{\dum}}
\newcommand{\phidum}{\varphi_{\dum}}

\newcommand{\om}{\omega}
\newcommand{\omb}{\overline{\omega}}

\newcommand{\Ukappa}{U_{\kappa}}
\newcommand{\Ugamma}{U_{\gamma}}
\newcommand{\Ugammaone}{U_{\gamma_1}}
\newcommand{\UMap}{U_{\Map}}
\newcommand{\X}{\hat{X}}
\newcommand{\UX}{U_{\X}}
\newcommand{\WX}{W_{\X}}
\newcommand{\Wkappa}{W_{\kappa}}
\newcommand{\Wgamma}{W_{\gamma}}
\newcommand{\Wgammaone}{W_{\gamma_1}}
\newcommand{\WMap}{W_{\Map}}
\newcommand{\phiX}{\varphi_{\X}}
\newcommand{\Maph}{\hat{\Map}}
\newcommand{\phiMap}{\varphi_{\Maph}}
\newcommand{\IX}{I_{\X}}
\newcommand{\yspMap}{y_{s,\Maph}^+}
\newcommand{\ysmMap}{y_{s,\Maph}^-}
\newcommand{\IMap}{I_{\Map}}

\title[Statistics of weak lensing shear and aperture-mass]
{Weak lensing shear and aperture-mass from linear to non-linear scales}
\author[Munshi D. et al.]
{Dipak Munshi$^{1,2}$, Patrick Valageas$^{3}$, Andrew J. Barber$^{4}$\\
$^{1}$Institute of Astronomy, Madingley Road,
Cambridge, CB3 OHA, United Kingdom\\
$^{2}$Astrophysics Group, Cavendish Laboratory, Madingley Road, 
Cambridge CB3 OHE, United Kingdom\\
$^{3}$ Service de Physique Th\'eorique, 
CEA Saclay, 91191 Gif-sur-Yvette, France \\
$^{4}$Astronomy Centre, University of Sussex, Falmer, Brighton, BN1 9QJ,
United Kingdom\\
}

\begin{document}
\maketitle

\begin{abstract}

In this paper we describe the predictions for the smoothed weak
lensing shear and aperture-mass of two simple analytical models of the
density field: the minimal tree-model and the stellar model. Both
models give identical results for the statistics of the 3-d density
contrast smoothed over spherical cells and only differ by the detailed
angular dependence of the many-body density correlations. We have
shown in previous work that they also yield almost identical results
for the pdf of the smoothed convergence, $\kappa_s$. We find that both
models give rather close results for both the shear and the positive
tail of the aperture-mass. However, we note that at small angular
scales ($\theta_s \la 2'$) the tail of the pdf $\cP(\Map)$ for
negative $\Map$ shows a strong variation between the two models and
the stellar model actually breaks down for $\theta_s \la 0.4'$ and
$\Map<0$. This shows that the statistics of the aperture-mass provides
a very precise probe of the detailed structure of the density field,
as it is sensitive to both the amplitude and the detailed angular
behaviour of the many-body correlations. On the other hand, the
minimal tree-model shows good agreement with numerical simulations
over all scales and redshifts of interest, while both models provide a
good description of the pdf $\cP(\gamma_{is})$ of the smoothed shear
components. Therefore, the shear and the aperture-mass provide robust
and complimentary tools to measure the cosmological parameters as well
as the detailed statistical properties of the density field.

\end{abstract}

 

\section{Introduction}

Magnification and shearing in the images of high-redshift galaxies
arise naturally due to gravitational lensing effects. These effects
derive from the fluctuations in the gravitational potential related to
the underlying density field. Statistical analysis of such observed
weak lensing data is therefore very effective in probing the
underlying density, which is assumed to be dominated by the dark
matter content. Observational surveys (see, e.g., Bacon, Refregier \&
Ellis, 2000, Hoekstra et al., 2002, Van Waerbeke et al., 2000, and Van
Waerbeke et al., 2002) have consequently been particularly fruitful in
this regard and have enabled estimates for the cosmological parameters
to be made.

To model weak gravitational lensing, cosmological $N$-body simulations
have been widely used in which the particle and mass distributions
have been set to reflect those expected in the real universe. The
first numerical studies for gravitational lensing employing $N$-body
simulations used ray-tracing methods (see, e.g., Schneider \& Weiss,
1988, Jarosszn'ski et al., 1990, Wambsganns, Cen \& Ostriker, 1998,
Van Waerbeke, Bernardeau \& Mellier, 1999, and Jain, Seljak \& White,
2000) to follow the deflections of light from sources at high redshift
in the simulations. More recently, Couchman, Barber \& Thomas (1999)
developed an algorithm to compute the full 3-dimensional shear
matrices at locations along lines of sight throughout simulation
volumes. Barber (2002) extended the method to combine the shear
matrices in the appropriate fashion along the lines of sight to
produce final Jacobian matrices from which the weak lensing statistics
could be directly derived.

On large angular scales, analytical computations for weak lensing
statistics can readily be made, as this is the regime where
perturbative calculations apply (e.g., Villumsen, 1996, Stebbins,
1996, Bernardeau et al., 1997, Jain \& Seljak, 1997, Kaiser, 1998, Van
Waerbeke, Bernardeau \& Mellier, 1999, and Schneider et al.,
1998). However, on small angular scales, especially relevant to
observational surveys with small sky coverage, perturbative
calculations are no longer valid and models to represent the
gravitational clustering in the non-linear regime have been devised.

To describe the non-linear evolution of the matter power spectrum,
Hamilton et al. (1991) proposed a technique based on a non-local
transformation. Their method was extended by Peacock \& Dodds (1996)
and included the conservation of mass and the rescaling of physical
lengths in the different regimes, following the ``stable-clustering''
{\it Ansatz} of Peebles (1980). More recently, Peacock \& Smith (2000)
and Seljak (2000) developed the ``Halo Model'' for the non-linear
evolution. This model is able to reproduce the matter power spectrum
of $N$-body simulations over a wide range of scales and relates the
linear and non-linear power at the same scale through fitting
formul\ae. In a more recent development, Smith et al. (2002) have
presented a new set of fitting functions based on the Halo Model and
calibrated to a set of $N$-body simulations.

Barber \& Taylor (2003) have now shown excellent agreement between the
power spectrum in the lensing convergence obtained from numerical
simulations in which they calculated the full 3-dimensional shear
matrices along lines of sight and the predictions from the Halo model
fitting functions of Smith et al. (2002).

An alternative approach used to describe the probability distribution
function of the density field, from which the associated weak lensing
statistics can be obtained, has been based directly on the many-body
correlations. The most common model of this kind expresses the
$p-$point correlations as a sum of $p-1$ products over the two-point
correlations linking all $p$ points. This yields the class of
``Hierarchical models,'' which are specified by the weights given to
any such topology associated with the $p-1$ products (e.g., Fry 1984,
Schaeffer 1984, Bernardeau \& Schaeffer 1992, and Szapudi \& Szalay
1993, 1997, Munshi et al. 1999a, Munshi, Melott, Coles 1999d). Once 
these weights have been assigned it is possible to
resum all many-body correlations and to compute the pdf of the density
field, or of any quantity which is linearly dependent on the matter
density. This is most easily done for ``minimal tree-models'', where
the weight associated to a given tree-topology is set by its vertices
(e.g., Bernardeau \& Schaeffer 1992, Munshi, Coles \& Melott 1999b,
Munshi, Coles \& Melott 1999d), or for ``stellar models'', which
only contain stellar diagrams (Valageas, Barber \& Munshi, 2003).

Using such an approach, coupled to the Hamilton et al. (1991)
prescription for the two-point correlation, Valageas (2000a, b) and
Munshi \& Jain (2000 and 2001), Munshi \& Coles (2000,2002) were able 
to compute the pdf of the weak-lensing convergence whilst the 
associated bias was considered by
Munshi (2000) and the cumulant correlators associated with such
distributions were evaluated by Munshi \& Jain (2000). Munshi \& Wang
(2003) further extended these studies to cosmological scenarios
including dark energy. These methods can also handle more intricate
quantities like the aperture-mass or the shear which involve
compensated filters and require a detailed model for the many-body
correlations. Thus, using a minimal tree-model for the non-linear
regime, Bernardeau and Valageas (2000) were able to predict the pdf of
the aperture-mass and to obtain a good agreement with numerical
simulations (they also showed that similar techniques could be applied
to the quasi-linear regime where the calculations can actually be made
rigorous). On the other hand, adopting a stellar model for the
many-body correlations, Valageas, Barber \& Munshi (2004) obtained
excellent agreement for the shear pdf when compared with the results
of $N$-body simulations. In a more recent paper, Barber, Munshi \&
Valageas (2004) described the results for the pdf of the convergence
based on hierarchical models and again showed excellent agreement with
the results from $N$-body simulations.

%
%
In this paper, we employ the minimal-tree model and the stellar model
for the density field to predict statistics for the weak lensing shear
and aperture mass, which correspond to a top-hat smoothing filter and a 
compensated filter, respectively. While earlier studies (e.g., Bernardeau \& 
Valageas 2000; Valageas, Barber \& Munshi 2004) showed that such approaches
provide a promising tool to obtain quantitative predictions for weak-lensing
observables, we compare in details in this article these theoretical 
predictions against results from numerical simulations, over a large range
of scales and redshifts. Thus, we are able to check that these methods
provide indeed reliable means to predict weak lensing statistics, from 
quasi-linear scales up to highly non-linear scales. In addition,
we compare the predictions obtained from the minimal-tree model and the 
stellar model. This allows us to investigate the sensitivity of these
weak-lensing observables onto the detailed angular behaviour of the many-body
density correlations (their overall amplitude at a given scale being identical
for both models). We find that such a dependence only shows up in the
negative tail of the aperture-mass at non-linear scales (which can then be 
used to discriminate between different angular models while other regimes
set the amplitude of the density correlations). Therefore, this paper
extends to the weak lensing shear and aperture mass the detailed analysis
presented in Barber et al.(2004) for the convergence (which is somewhat
simpler).
%
%

In section~2, we describe weak lensing
distortions in general and the use of filters to smooth the data. In
section~3, we introduce the minimal-tree and stellar models for the
density field and in section~4, we describe how the pdfs for the weak
lensing observables are derived in these models. In section~5, we
describe the $N$-body simulations and the procedure for computing the
lensing statistics in the simulations; we also outline the procedure
for binning the data and applying the smoothing filters. In section~6,
we declare the results in detail which are discussed finally in
section~7.

\section{Weak lensing distortions}
\label{Weak lensing distortions}

As a photon travels from a distant source towards the observer, its 
trajectory is deflected by density fluctuations close to the line-of-sight. 
This leads to an apparent displacement of the source and to a distortion of
the image (as the deflection varies with the direction on the sky). These 
effects can be observed through the amplification or the shear of the observed
images of distant sources. One such measure of the distortion may be
given in terms of the convergence 
along the line-of-sight, $\kappa({\vec \vartheta})$, given by (e.g., 
Bernardeau et al., 1997; Kaiser, 1998):
\beq
\kappa({\vec \vartheta}) = \frac{3\Om}{2} \int_0^{\chi_s} \d\chi \; 
w(\chi,\chi_s) \; \delta(\chi,\De{\vec \vartheta}) ,
\label{kappa}
\eeq
with:
\beq
w(\chi,\chi_s) = \frac{H_0^2}{c^2} \; \frac{\De(\chi) \De(\chi_s-\chi)}
{\De(\chi_s)} \; (1+z) ,
\label{w}
\eeq
where $z$ corresponds to the radial distance $\chi$ and $\De$ is the angular 
distance. Here and in the following we use the Born approximation which is
well-suited to weak-lensing studies: the fluctuations of the gravitational
potential are computed along the unperturbed trajectory of the photon 
(Kaiser, 1992).
Thus the convergence $\kappa({\vec \vartheta})$ is merely the 
projection 
of the local density contrast $\delta$ along the line of sight up to the 
redshift $z_s$ of the source. Therefore, weak lensing observations allow us to
measure the projected density field $\kappa({\vec \vartheta})$ on the sky. 
Note that by 
looking at sources located at different redshifts one may also probe the radial
direction. From eq.(\ref{kappa}) we can see that there is a 
minimum value, $\kappamin(z_s)$, for the convergence of a source located at 
redshift $z_s$, which corresponds to an ``empty'' beam between the source 
and the observer ($\delta=-1$ everywhere along the line of sight):
\beq
\kappamin = - \frac{3\Om}{2} \int_0^{\chi_s} \d\chi \; w(\chi,\chi_s) .
\label{kappamin}
\eeq
Following Valageas (2000a, b) it is convenient to define the ``normalized'' 
convergence, $\kappah$, by:
\beq
\kappah = \frac{\kappa}{|\kappamin|} = \int_0^{\chi_s} \d\chi \; \wh \; 
\delta , \hspace{0.2cm} \mbox{with} \hspace{0.2cm} 
\wh=\frac{w(\chi,\chi_s)}{\int_0^{\chi_s} \d\chi \; w(\chi,\chi_s)} ,
\label{kappah}
\eeq
which obeys $\kappah \geq -1$. Here we introduced the ``normalized selection 
function,'' $\wh(\chi,\chi_s)$. 

In practice, one does not study the projected density 
$\kappa({\vec \vartheta})$
itself but applies first a smoothing procedure. For instance, it is customary
to investigate the ``smoothed convergence'' $\kaphs$ (where the subscript 
``s'' refers to ``smoothed'') given by:
\beq
\kaphs = \int \d{\vec \vartheta} \; \Ukappa({\vec \vartheta}) \; 
\kappah({\vec \vartheta})
\label{kapthe} ,
\eeq
with:
\beq
\Ukappa({\vec \vartheta}) = \frac{\Theta(\vartheta<\theta_s)}{\pi\theta_s^2},
\label{Ukappa}
\eeq
where $\Theta(\vartheta<\theta_s)$ is a top-hat with obvious notations. 
Thus, the ``smoothed convergence'' $\kaphs$ is simply the projected density
contrast $\kappa({\vec \vartheta})$ smoothed with a normalised top-hat 
$\Ukappa$ of angular radius $\theta_s$. By varying the radius $\theta_s$ one 
probes the density field at different wavelengths.

The ``smoothed convergence'' $\kaphs$ was already studied at length in 
previous papers (e.g., Barber, Munshi \& Valageas 2004). Here we focus on the shear 
$\gamma$ and the aperture mass $\Map$. These two quantities can again be 
expressed in terms of the projected density contrast 
$\kappa({\vec \vartheta})$ as in eq.(\ref{kapthe}). Thus, the smoothed shear 
$\gamma_s=\gamma_{1s}+i\gamma_{2s}$ corresponds to the filter $\Ugamma$ 
given by (see Valageas, Barber \& Munshi 2003):
\beq
\Ugamma({\vec \vartheta}) = - \frac{\Theta(\vartheta>\theta_s)}
{\pi\vartheta^2} \; e^{i2\alpha} ,
\label{Ugamma}
\eeq
where $\Theta(\vartheta>\theta_s)$ is again a Heaviside function with obvious
notations and $\alpha$ is the polar angle of the vector ${\vec \vartheta}$ 
with the 1-axis (the 2-axis has $\alpha=\pi/2$). The comparison of 
eq.(\ref{Ugamma}) with eq.(\ref{Ukappa}) shows that while the 
smoothed convergence only depends on the matter {\it within} the cone formed 
by the angular window, $\theta_s$, the smoothed shear only depends on the 
matter {\it outside} this cone. Then, the smoothed shear component $\gamma_1$
along the 1-axis is described by the filter $\Ugammaone$:
\beq
\Ugammaone({\vec \vartheta}) = - \frac{\Theta(\vartheta>\theta_s)}
{\pi\vartheta^2} \; \cos 2\alpha .
\label{Ugamma1}
\eeq
Note that the filters $\Ugamma$ and $\Ugammaone$ depend on both the length and
the angle of the two-dimensional vector ${\vec \vartheta}$ in the plane 
perpendicular to the mean line-of-sight (in this article we only consider 
small angular scales). Moreover, they are compensated filters which is an 
interesting feature since convergence maps are only reconstructed up to
a mass sheet degeneracy. In fact, the shear $\gamma$ is actually the quantity
which is most directly linked to observations.

One drawback of the shear components is that they are even quantities (their
sign can be changed through a rotation of axis, see eq.(\ref{Ugamma})) hence
their third-order moment vanishes by symmetry and one must measure the 
fourth-order moment $\lag \gamma_1^4\rag$ (i.e. the kurtosis) in order to
probe the deviations from Gaussianity. Another quantity which involves a
compensated filter but is not even is the aperture-mass $\Map$. It simply
corresponds to a compensated filter with polar symmetry: 
$\UMap({\vec \vartheta}) = \UMap(\vartheta)$. For instance, following
Schneider (1996) one can use:
\beq
\UMap({\vec \vartheta}) = \frac{\Theta(\vartheta<\theta_s)}{\pi\theta_s^2}
\; 9 \left(1-\frac{\vartheta^2}{\theta_s^2}\right) 
\left(\frac{1}{3} - \frac{\vartheta^2}{\theta_s^2}\right) .
\label{UMap}
\eeq
The advantage of such compensated filters is that one can also express $\Map$
as a function of the tangential component $\gamma_t$ of the shear (Kaiser et
al. 1994; Schneider 1996) so that it is not necessary to build a full 
convergence map from observations. Besides, the aperture-mass provides a
useful separation between $E$ and $B$ modes.

In the following, we shall write $\UX$ for arbitrary filters when our results
apply to any quantity $\X$ defined as in eq.(\ref{kapthe}) with some filter 
$\UX$, and we shall specify $\Ukappa,..,\UMap$ for particular 
cases. For some purposes it is convenient to work in Fourier space, therefore
we define the Fourier transform of the density contrast by:
\beq
\delta(\bx) = \int \d\bk \; e^{i \bk.\bx} \; \delta(\bk),
\label{deltak}
\eeq
where $\bx$ and $\bk$ are comoving coordinates. Then, we define the 
power-spectrum $P(k)$ of the density contrast by:
\beq
\lag \delta(\bk_1) \delta(\bk_2) \rag = \delta_D(\bk_1+\bk_2) \; P(k_1),
\label{Pk}
\eeq
where $\delta_D$ is Dirac's distribution. This yields for the two-point
correlation $\xi_2(x)$ of the density contrast:
\beq
\xi_2(x) = \lag \delta(\bx_1) \delta(\bx_1+\bx) \rag = \int \d\bk \; 
e^{i\bk.\bx} \; P(k) .
\label{xi2}
\eeq
We also introduce the power per logarithmic interval $\Delta^2(k)$ as:
\beq
\Delta^2(k,z) = 4 \pi k^3 P(k,z) .
\label{Delta2}
\eeq
Then, we can write for any normalised quantity $\X$ (such as the normalised 
smoothed shear $\gamhs$):
\beqa
\lefteqn{\X = \int \d{\vec \vartheta} \; \UX({\vec \vartheta}) \; 
\kappah({\vec \vartheta}) = \int\d\chi \; \wh \int \d{\vec \vartheta}
\; \UX({\vec \vartheta}) \delta(\chi,\De {\vec \vartheta}) }  \label{Xx} \\
& & = \int_0^{\chi_s}\d\chi \; \wh \; \int\d\bk \;
e^{i\kpar\chi} \; \WX(\kperp\De\theta_s) \; \delta(\bk) ,
\label{Xk}
\eeqa
where $\kpar$ is the component of $\bk$ parallel to the line-of-sight,
$\kperp$ is the two-dimensional vector formed by the components of $\bk$
perpendicular to the line-of-sight and
 $\WX$ is the Fourier form of the real-space filter $\UX$:
\beq
\WX(\kperp\De\theta_s) = \int\d{\vec \vartheta} \; \UX({\vec \vartheta}) 
\; e^{i \kperp.\De{\vec \vartheta}} .
\label{WX}
\eeq
Following previous works we explicitly introduced the angular scale 
$\theta_s$ in the definition of the Fourier filter $\WX$. For the smoothed
convergence $\kaphs$ we obtain:
\beq
\Wkappa(\kperp\De\theta_s) = \frac{2 J_1(\kperpDt)}{\kperpDt} ,
\label{Wkappa}
\eeq
where $J_1$ is the Bessel function of the first kind of order 1, while for
the shear we have:
\beq
\Wgamma(\kperp\De\theta_s) = \Wkappa(\kperpDt) \; e^{i2\alpha}
\label{Wgamma}
\eeq
and:
\beq
\Wgammaone(\kperp\De\theta_s) = \Wkappa(\kperpDt) \; \cos2\alpha ,
\label{Wgammaone}
\eeq
where $\alpha$ is the polar angle of the transverse wavenumber $\kperp$.
Finally, we obtain for the aperture-mass defined from the filter 
(\ref{UMap}):
\beq
\WMap(\kperp\De\theta_s) = \frac{24 J_4(\kperpDt)}{(\kperpDt)^2} .
\label{WMap}
\eeq
In any case, the variance of the observable $\X$ is obtained from 
eq.(\ref{Xk}) as:
\beq
\lag\X^2\rag_c = \int_0^{\chi_s} \d\chi \; \wh^2 \; \frac{1}{2} 
\int \frac{\d\kperp}{k_{\perp}^2} \; \frac{\Delta^2(k_{\perp})}{k_{\perp}}
\; \WX(\kperp\De\theta_s)^2 .
\label{varX}
\eeq

\section{The density field}
\label{The density field}

In order to derive the properties of weak lensing observables, like the
shear $\gamma$, we clearly need to specify the properties of the 
underlying density field. In this paper we use simple models which we
described in detail in Valageas, Barber \& Munshi (2003) and Barber, Munshi \&
Valageas (2003), 
hence we only briefly recall
the main elements of this framework here. First, we describe the pdf 
$\cP(\delta_R)$ of the density contrast at scale $R$ through its generating 
function $\varphi(y)$:
\beq
e^{-\varphi(y)/\xib_2} = \int_{-1}^{\infty} \d\delta_R \; 
e^{-\delta_R y/\xib_2} \; \cP(\delta_R) ,
\label{phidelta}
\eeq
where $\delta_R$ is the density contrast within spherical cells of radius $R$
and volume $V$ while $\xib_2$ is its variance:
\beq
\delta_R = \int_V \frac{\d\bx}{V} \; \delta(\bx) \hspace{0.3cm} \mbox{and}
\hspace{0.3cm} \xib_2 = \lag \delta_R^2 \rag .
\label{deltaR}
\eeq
The function $\varphi(y)$ defined from eq.(\ref{phidelta})
is also the generating function of the cumulants of the density contrast
and its expansion at $y=0$ reads:
\beq
\varphi(y) = \sum_{p=2}^{\infty} \frac{(-1)^{p-1}}{p!} \; S_p \; y^p 
\hspace{0.3cm} \mbox{with} \hspace{0.3cm} 
S_p = \frac{\lag\delta_R^p\rag_c}{\xib_2^{\; p-1}} .
\label{phiSp}
\eeq
As described in Valageas, Barber \& Munshi (2003) and Barber, Munshi \& Valageas (2003) we 
parameterize the generating function $\varphi(y)$ through the skewness $S_3$
of the density field, which we estimate as:
\beq
S_3(z)= S_3^{\rm QL} + \frac{\Delta^2(k_s,z) -1}{\Delta_{\rm vir}(z)-1} 
\left( S_3^{\rm NL} - S_3^{\rm QL} \right) .
\label{S3}
\eeq
Here, $S_3^{\rm QL}$ is the exact result derived in the quasi-linear limit
while $S_3^{\rm NL}$ is the prediction of HEPT 
(Scoccimarro \& Frieman 1999) for the highly non-linear regime. We also 
introduced the density contrast $\Delta_{\rm vir}$ at virialization which
marks the onset of the highly non-linear regime. Eq.(\ref{S3}) applies to
intermediate scales. At large scales ($\Delta^2(k_s,z)<1$) we simply take
$S_3= S_3^{\rm QL}$ while at small scales ($\Delta^2(k_s,z)>\Delta_{\rm vir}$)
we use $S_3= S_3^{\rm NL}$. Finally, we define the typical wavenumber $k_s(z)$
probed by weak lensing observations as:
\beq
k_s(z) = \frac{1}{\De(z)\theta_s} \hspace{0.2cm} \mbox{for $\kappa$ or $\gamma$;} \hspace{0.2cm} k_s(z) = \frac{4}{\De(z)\theta_s} \hspace{0.2cm} \mbox{for $\Map$} .
\label{ks}
\eeq
Indeed, as seen from Fig.~\ref{FigD2k}, for the same angular radius 
$\theta_s$ the aperture-mass $\Map$ probes higher wavenumbers than the 
convergence or the shear. This is obviously due to the radial structure of 
the filter $\UMap(\vartheta)$. As seen from eq.(\ref{phidelta}), 
in order to determine the pdf 
$\cP(\delta_R)$ we simply need to add a prescription for the variance 
$\xib_2$. We use the fit to numerical simulations provided by Peacock \&
Dodds (1996).

>From the pdf $\cP(\delta_R)$, or the generating function $\varphi(y)$ and
the variance $\xib_2$, we have a full description of the density fluctuations
smoothed over spherical cells. As shown in detail in Valageas (2000) and 
Barber, Munshi \& Valageas (2003) 
this is sufficient to obtain the properties of the smoothed convergence
$\kappa_s$. However, as noticed in Bernardeau \& Valageas (2000) and Valageas
et al. (2003), this information is no longer sufficient when we study more
intricate observables like the shear or the aperture-mass which involve
compensated filters. Therefore, we need to specify the detailed angular
behaviour of the many-body correlation functions $\xi_p(\bx_1,..,\bx_p)$, 
defined by (Peebles 1980):
\beq
\xi_p (\bx_1,..,\bx_p) = \lag \delta(\bx_1) .. \delta(\bx_p) \rag_c .
\label{xip}
\eeq
As in Barber, Munshi \& Valageas (2003) we shall compare two specific cases within the more
general class of ``tree-models''. The latter are defined by the 
hierarchical property (Schaeffer, 1984, and Groth \& Peebles, 1977):
\beq
\xi_p(\bx_1, .. ,\bx_p) = \sum_{(\alpha)} Q_p^{(\alpha)} \sum_{t_{\alpha}} 
\prod_{p-1} \xi_2(\bx_i,\bx_j) 
\label{tree}
\eeq
where $(\alpha)$ is a particular tree-topology connecting the $p$ points 
without making any loop, $Q_p^{(\alpha)}$ is a parameter associated with the 
order of the correlations and the topology involved, $t_{\alpha}$ is a 
particular labeling of the topology, $(\alpha)$, and the product is made over 
the $(p-1)$ links between the $p$ points with two-body correlation functions. 

Then, the ``minimal tree-model'' corresponds to the specific case where the 
weights $Q_p^{(\alpha)}$ are given by (Bernardeau \& Schaeffer, 1992): 
\beq
Q_p^{(\alpha)} = \prod_{\mbox{vertices of } (\alpha)} \nu_q
\label{mintree}
\eeq
where $\nu_q$ is a constant weight associated to a vertex of the tree 
topology with $q$ outgoing lines. The advantage of this minimal tree-model
is that it is well-suited to the computation of the cumulant generating
functions as defined in eq.(\ref{phiSp}) for the density contrast 
$\delta_R$. Indeed, for an arbitrary real-space filter, $F(\bx)$, which 
defines the random variable $\dum$ as:
\beq
\dum = \int \d\bx \; F(\bx) \; \delta(\bx) \hspace{0.4cm} \mbox{and} 
\hspace{0.4cm} \xidum= \lag \dum^2 \rag ,
\label{mu}
\eeq
it is possible to obtain a simple implicit expression for the 
generating function, $\phidum(y)$ (see Bernardeau \& Schaeffer, 1992,
and Jannink \& Des Cloiseaux, 1987): 
\beqa
{\displaystyle \phidum(y)} & = & {\displaystyle y \int \d\bx \; F(\bx) \; 
\left[ \zeta_{\nu}[ \tau(\bx)] - \frac{\tau(\bx) \zeta'_{\nu}[\tau(\bx)]}{2} 
\right] } 
\label{implicitphi} \\ 
{\displaystyle \tau(\bx) } & = & {\displaystyle -y \int \d\bx' \; F(\bx') 
\; \frac{\xi_2(\bx,\bx')}{\xidum} \; \zeta'_{\nu}[\tau(\bx')] } 
\label{implicittau}
\eeqa
where the function $\zeta_{\nu}(\tau)$ is defined as the generating function 
for the coefficients $\nu_p$:
\beq
\zeta_{\nu}(\tau) = \sum_{p=1}^{\infty} \frac{(-1)^p}{p!} \; \nu_p \; \tau^p 
\hspace{0.4cm} \mbox{with} \hspace{0.4cm} \nu_1 = 1.  
\label{zetanu}
\eeq

A second simple model is the ``stellar model'' introduced in Valageas, Barber \& Munshi (2003) where we only keep the stellar diagrams in eq.(\ref{tree}). Thus, 
the $p-$point connected correlation $\xi_p$ of the density field can now 
be written as:
\beq
\xi_p(\bx_1, .. ,\bx_p) = \frac{\tS_p}{p} \; \sum_{i=1}^p \prod_{j\neq i} 
\xi_2(\bx_i,\bx_j) .
\label{stellar}
\eeq
The advantage of the stellar-model (\ref{stellar}) is that it leads to 
very simple calculations in Fourier space. Indeed, eq.(\ref{stellar}) reads 
in Fourier space:
\beq
\lag \delta(\bk_1) .. \delta(\bk_p) \rag_c = \frac{\tS_p}{p} \; 
\delta_D(\bk_1+..+\bk_p) \sum_{i=1}^p \prod_{j\neq i} P(k_j) .
\label{stellark}
\eeq
Following Valageas, Barber \& Munshi (2003) and Barber, Munshi \& Valageas (2003) we shall use the 
simple approximation $\tS_p \simeq S_p$. Alternatively, we may define the 
function $\varphi(y)$ as the generating function of the coefficients 
$\tS_p$, rather than $S_p$, through its Taylor expansion at $y=0$.

\section{The pdf of weak lensing observables}
\label{The pdf of weak lensing observables}

>From the models described in Sect.~\ref{The density field} for the density
field we can derive the pdf of the weak lensing observables defined in
Sect.~\ref{Weak lensing distortions}. The procedure is described in detail
in Barber, Munshi \& Valageas (2003) (and references therein). Hence we only briefly 
recall here the main results.

\subsection{Minimal tree-model}
\label{Minimal tree-model}

We first apply the minimal tree-model to an arbitrary weak-lensing 
observable $\X$ defined by the filter $\UX$. As described in Valageas
(2000b), in order to derive the pdf $\cP(\X)$ we first compute the
cumulants $\lag\X^p\rag_c$, which we resum to obtain the generating function
$\phiX(y)$ defined as in eq.(\ref{phiSp}):
\beq
\phiX(y)= \sum_{p=1}^{\infty} \frac{(-1)^{p-1}}{p!} \; 
\frac{\lag\X^p\rag_c}{\lag\X^2\rag_c^{p-1}} \; y^p .
\label{phiX}
\eeq
Next, the pdf $\cP(\X)$ is given by the inverse Laplace transform:
\beq
\cP(\X) = \inta \frac{\d y}{2\pi i \lag\X^2\rag_c} \; 
e^{[\X y - \phiX(y)] / \lag\X^2\rag_c} .
\label{PX}
\eeq
Thus, we first obtain the cumulants $\lag\X^p\rag_c$ from (\ref{Xx}):
\beqa
\lag \X^p\rag_c & = & \int_0^{\chi_s} \d\chi \; \wh^p 
\int_{-\infty}^{\infty} \prod_{i=2}^{p} \d\chi_i \int \prod_{i=1}^{p} 
\d{\vec \vartheta}_i \; \UX({\vec \vartheta}_i) \nonumber \\ 
& & \times \; \xi_p\left( \bea{l} 0 \\ \De {\vec \vartheta}_1 \ea ,
\bea{l} \chi_2 \\ \De {\vec \vartheta}_2 \ea , .. , 
\bea{l} \chi_p \\ \De {\vec \vartheta}_p \ea ; z \right) .
\label{cumX}
\eeqa
Then, as seen in Valageas (2000b) we note that within the framework of 
a tree-model (\ref{tree}) the 2-d correlations $\om_p$ exhibit the same 
tree-structure, with:
\beqa
\om_p({\vec \vartheta}_1,..,{\vec \vartheta}_p;z) & = & 
\int_{-\infty}^{\infty} \prod_{i=2}^{p} \d\chi_i \; 
\xi_p\left( \bea{l} 0 \\ \De {\vec \vartheta}_1 \ea , .. ,
\bea{l} \chi_p \\ \De {\vec \vartheta}_p \ea ; z \right) \nonumber \\
\label{omp}
\eeqa
In terms of the power-spectrum $P(k)$ we can also write $\om_2$ as:
\beq
\om_2({\vec \vartheta}_1,{\vec \vartheta}_2;z) = \pi \int_0^{\infty} 
\frac{\d k}{k} \; \frac{\Delta^2(k,z)}{k} \; 
J_0(k \De |{\vec \vartheta}_1 - {\vec \vartheta}_2|) .
\label{om2}
\eeq
Then, in the case of a minimal tree-model (\ref{mintree}) we can perform 
the resummation (\ref{implicitphi})-(\ref{implicittau}) for the 2-d 
correlations $\om_p$, since the latter obey the same minimal tree-model. 
This yields (see Bernardeau \& Valageas 2000 and Barber, Munshi \& Valageas 2003 
for details):
\beq
\phiX(y) = \int_0^{\chi_s} \d\chi \; \frac{\lag\X^2\rag_c}{\omb_{2\X}} \;
\varphi_{\rm cyl.}\left( y \wh \frac{\omb_{2\X}}{\lag\X^2\rag_c} ; z \right) ,
\label{phiXtree}
\eeq
where we introduced the 2-d generating function $\varphi_{\rm cyl.}$ 
associated with the 2-d correlations $\om_p$, given by the resummation:
\beqa
{\displaystyle \varphi_{\rm cyl.}(y)} & = & {\displaystyle y 
\int \d{\vec \vartheta} \; \UX({\vec \vartheta}) \; 
\left[ \zeta_{\nu}[ \tau({\vec \vartheta})] 
- \frac{\tau({\vec \vartheta}) \zeta'_{\nu}[\tau({\vec \vartheta})]}{2} 
\right] } 
\label{phiom} \\ 
{\displaystyle \tau({\vec \vartheta}) } & = & {\displaystyle -y 
\int \d{\vec \vartheta}' \; \UX({\vec \vartheta}') \;
\frac{\om_2({\vec \vartheta},{\vec \vartheta}';z)}{\omb_{2\X}(z)} \; 
\zeta'_{\nu}[\tau({\vec \vartheta}')] } 
\label{tauom}
\eeqa
Here we introduced the angular average $\omb_{2\X}$ of the 2-d correlation 
$\om_2$, associated with the filter $\UX$:
\beq
\omb_{2\X}(z) = \int \d{\vec \vartheta}_1 \d{\vec \vartheta}_2 \;
\UX({\vec \vartheta}_1) \UX({\vec \vartheta}_2) \;
\om_2({\vec \vartheta}_1,{\vec \vartheta}_2;z) .
\label{omb2X}
\eeq
Thus, we obtain in this way both generating functions $\phigamone$ and
$\phiMap$ associated with the smoothed normalised shear component 
$\gamonehs$ and the aperture-mass $\Maph$. This yields in turn the pdfs
$\cP(\gamonehs)$ and $\cP(\Maph)$, using eq.(\ref{PX}). Note that for 
the aperture-mass the implicit system (\ref{phiom})-(\ref{tauom})
simplifies somewhat since the filter $\UMap({\vec \vartheta})$ only 
depends on the length $\vartheta$.

\subsection{Stellar model}
\label{Stellar model}

Next, we can use the same procedure within the framework of the stellar 
model (\ref{stellar}). Thus, working in Fourier space we obtain from 
eq.(\ref{Xk}) the cumulants $\lag\X^p\rag_c$ as:
\beqa
\lag \X^p \rag_c & = & \int \frac{\d\chi}{2\pi} (2\pi\wh)^p 
\int \prod_{j=1}^{p} \d\bk_{\perp j} \; \WX(\bk_{\perp j} \De \theta_s) 
\nonumber \\ 
& & \times \;\tS_p \; \delta_D(\bk_{\perp 1}+..+\bk_{\perp p}) 
P(k_{\perp 2}) .. P(k_{\perp p}) .
\label{cumXk1}
\eeqa
Next, using the standard exponential representation of the Dirac 
distribution (see Valageas, Barber \& Munshi 2003) and eq.(\ref{WX}), we 
can write:
\beq
\lag \X^p \rag_c = \int_0^{\chi_s} \d\chi \; \wh^p \int \d{\vec \vartheta}
\; \UX({\vec \vartheta}) \; \tS_p \; \IX(\chi,{\vec \vartheta})^{p-1} ,
\label{cumXk2}
\eeq
where we introduced:
\beq
\IX(\chi,{\vec \vartheta}) = \frac{1}{2} \int 
\frac{\d\bk_{\perp}}{k_{\perp}^2} \; \frac{\Delta^2(k_{\perp},z)}{k_{\perp}} 
\; \WX(\bk_{\perp} \De \theta_s) \; e^{i\bk_{\perp} . \De {\vec \vartheta}} .
\label{IX}
\eeq
Then, using eq.(\ref{phiSp}) we obtain:
\beqa
\!\! \phiX(y) = \int_0^{\chi_s} \d\chi \int \d{\vec \vartheta}
\; \UX({\vec \vartheta}) \frac{\lag\X^2\rag_c}{\IX(\chi,{\vec \vartheta})} 
\; \varphi\left( y \wh \frac{\IX}{\lag\X^2\rag_c} ; z \right) .
\label{phiXstellar}
\eeqa
The result (\ref{phiXstellar}) allows us to obtain both generating functions 
$\phigamone$ and $\phiMap$ associated with the smoothed shear component 
$\gamonehs$ and the aperture-mass $\Maph$. This again yields in turn the pdfs
$\cP(\gamonehs)$ and $\cP(\Maph)$, using eq.(\ref{PX}). For 
the aperture-mass eq.(\ref{phiXstellar}) also
simplifies somewhat since the filter $\UMap({\vec \vartheta})$ and 
$\IMap(\chi,{\vec \vartheta})$ only depend on the length $\vartheta$.

\subsection{Exponential tails}
\label{Exponential tails}

As described in Bernardeau \& Schaeffer (1992), the implicit system 
(\ref{phiom})-(\ref{tauom}) usually yields branch cuts along the real axis
for the generating function $\varphi_{\rm cyl.}(y)$. In fact, we actually 
define the generating function $\varphi(y)$ of the 3-d density contrast 
through a similar implicit system (see Valageas, Barber \& Munshi 2003 and Barber, Munshi \& Valageas 2003 for details), so that the function $\varphi(y)$ shows a branch
cut along the negative real axis for $y<y_s$, with:
\beq
y_s = - \frac{\kappa}{\kappa+2} \left( \frac{\kappa+1}{\kappa+2} 
\right)^{\kappa+1} \hspace{0.3cm} \mbox{with} \hspace{0.3cm} 
\kappa= \frac{3}{S_3-3} .
\label{ys}
\eeq
Here $S_3$ is the skewness of the density contrast at the scale and time
of interest. The singularity $y_s$ leads to an exponential tail for the
pdf $\cP(\delta_R)$ for large positive $\delta_R$. On the other hand, for
large $y$ the generating function $\varphi(y)$ shows a slow power-law
growth (see Bernardeau \& Schaeffer 1992):
\beq
y \rightarrow +\infty : \;\; \varphi(y) + y \sim y^{1-\om} 
\hspace{0.3cm} \mbox{with} \hspace{0.3cm} 0<\om<1 .
\label{omphi}
\eeq
This large $y$ behaviour leads to a strong cutoff at low densities and
$\cP(\delta_R)=0$ for $\delta_R<-1$ (this lower bound is set by 
the coefficient of the term linear over $y$ in eq.(\ref{omphi})) when
we can push the integration path to $+\infty$ in eq.(\ref{PX}).
As seen in Valageas (2000) and Barber, Munshi \& Valageas (2003), since the smoothed
convergence is described by a 2-d top-hat, which is quite similar to the
3-d top-hat associated with the density contrast $\delta_R$, the generating
functions $\varphi_{\rm cyl.}(y)$, $\phikap$ and the pdf $\cP(\kappa_s)$
show the same behaviour as for $\delta_R$. In fact, as shown in Valageas 
(2000) and Barber, Munshi \& Valageas (2003), one can directly use $\varphi(y)$ and
$\cP(\delta_R)$ to obtain up to a good accuracy the properties of the 
smoothed convergence.

Of course, for more intricate filters the properties of the generating 
function $\phiX(y)$ and of the pdf $\cP(\X)$ can exhibit very different
behaviours. In particular, for the shear component $\gamma_{1s}$ the 
generating 
function $\phigamone(y)$ and the pdf $\cP(\gamonehs)$ are now even. This
property is obviously preserved within both models used in this paper.
For the minimal tree-model, this appears in the implicit system 
(\ref{phiom})-(\ref{tauom}) through the factor $\cos 2\alpha$ of the filter
$\Ugammaone({\vec \vartheta})$ given in eq.(\ref{Ugamma1}), which clearly
implies that the function $\varphi_{\rm cyl.}$ is even. On the other hand,
for the stellar model this property shows up in eq.(\ref{phiXstellar}) because
the angular dependence of the filter $\Ugammaone({\vec \vartheta})$ leads to
$\Igamone(\chi,{\vec \vartheta}) \propto \cos 2\alpha$, see Valageas, Barber \& Munshi (2003) for details. Then, in both cases the generating function 
$\phigamone(y)$ shows two symmetric branch cuts along the real axis for
$y<-\ysgamone$ and $y>\ysgamone$ (with $\ysgamone>0$). For the stellar model,
this singularity can be expressed in terms of $y_s$ as (see Valageas et 
al. 2003):
\beq
\ysgamone = \min_{z,\vartheta} \left| y_s 
\frac{\lag\gamonehs^2\rag_c}{\Igamone\wh} \right| .
\label{ysgam1}
\eeq
Then, the pdf $\cP(\gamonehs)$ shows two symmetric exponential tails for
$\gamonehs \rightarrow \pm \infty$.

On the other hand, the aperture-mass $\Map$ can lead to a more intricate
behaviour. Since it involves a compensated filter like the shear component
$\gamma_{1s}$, we can expect extended tails both for large positive and 
negative
$\Map$. However, the generating function $\phiMap(y)$ and the pdf $\cP(\Map)$
are not even. From the shape of the filter (\ref{UMap}) we can actually expect
the fall-off in the pdf to be sharper for negative $\Map$. For the stellar 
model, the two branch cuts along the real axis of the generating function 
$\phiMap(y)$ are given by:
\beq
\ysmMap = \frac{y_s}{\max_{z,\vartheta} (\wh\IMap/\lag\Maph^2\rag_c)} ,
\hspace{0.4cm} \ysmMap <0 ,
\label{ysmMap}
\eeq
and:
\beq
\yspMap = \frac{y_s}{\min_{z,\vartheta} (\wh\IMap/\lag\Maph^2\rag_c)} ,
\hspace{0.4cm} \yspMap >0 .
\label{yspMap}
\eeq
We can check numerically that $|\ysmMap|<\yspMap$ which means that the cutoff
of $\cP(\Map)$ is indeed stronger for negative $\Map$. This property is also
verified by the tree-model.

Here, we must point out that such exponential tails for the various pdf
(associated with a branch cut along the negative real axis for $\varphi(y)$) 
are a mere consequence of the parameterization we use for $\varphi(y)$.
Indeed, as recalled above, the pdf $\cP(\kappa_s)$ of the smoothed convergence
closely follows the pdf $\cP(\delta_R)$ of the 3-d density contrast. 
Therefore, if we model $\cP(\delta_R)$ in such a way that it shows a stronger
cutoff than the exponential at large overdensities (such as a Gaussian)
it would translate into a similar behaviour for $\cP(\kappa_s)$ and the
generating functions $\varphi(y)$ and $\varphi_{\kappa}(y)$ would have
no branch cut. Such a behaviour would also apply to the shear or
the aperture-mass. Similarly, a cutoff which is smoother than exponential
for $\cP(\delta_R)$ would yield extended tails for weak lensing observables.
In the quasi-linear limit it is possible to derive the tails of the pdf 
$\cP(\delta_R)$ and to show that the high-density cutoff actually involves the
exponential of some power-law (see Valageas 2002a) but this discrepancy is
negligible in the range of interest (the far tail corresponds to very rare
events which are irrelevant for most practical purposes). In the non-linear
regime there is no rigorous derivation of the tails of the pdf $\cP(\delta_R)$
(however see Valageas 2002b for more details) but the simple phenomenological
model recalled in Sect.~\ref{The density field} appears to be sufficient,
as seen in Barber, Munshi \& Valageas (2003) from a comparison with numerical simulations
for the weak lensing convergence $\kappa_s$.

\subsection{The tail of $\cP(\Map)$ for negative $\Map$ at small angles: 
sensitivity onto the angular behaviour of the many-body correlations}
\label{The tail}

We can note that at small scales, 
$\theta_s \la 0.4'$, the factor $\IMap$ becomes positive for all $z$ and
$\vartheta$, which means that the branch cut for positive $y$ is repelled to
$+\infty$ within the stellar model. Then, the pdf $\cP(\Map)$ actually vanishes
for $\Map<0$. Since we always have $\lag\Map\rag=0$, this means that the pdf
$\cP(\Map)$ actually contains a singular part at the origin. This behaviour
appears when the power per logarithmic interval $\Delta^2(k)$ is flatter
than $k$. For CDM-like power-spectra this corresponds to small scales 
(i.e. to $\theta_s \la 0.4'$, see Fig.~\ref{FigD2k}). Note that this 
behaviour is {\it independent} of the coefficients $S_p$, that is of our 
parameterization for the generating function $\varphi(y)$ or the 
pdf $\cP(\delta_R)$. It is only due to
the angular structure implied by the stellar model and would remain unchanged
whatever the prescription used for the pdf $\cP(\delta_R)$ of the density 
contrast.
This suggests that no physical density field can be
exactly described by the stellar model (\ref{stellar}). On the other hand,
the minimal tree-model does not show such a breakdown at small angles and
it always yields a smooth exponential tail for large negative $\Map$.

This discrepancy between both models shows that the tail of the pdf 
$\cP(\Map)$ for negative $\Map$ is very sensitive to the detailed angular
properties of the density field. This is an interesting feature since it means
that one could extract in principles some useful information about the 
structure of the density field at small scales from the aperture-mass $\Map$.
Unfortunately, such small angular scales may be difficult to study in
actual observations. However, as seen in 
Sect.~\ref{Probability Distribution Function for Aperture-Mass} a large 
discrepancy between both models is already apparent for a smoothing radius 
$\theta_s=2'$ which may be within the reach of observations.

\subsection{Integration path}
\label{Integration path}

\begin{figure}
\begin{center}
\psfig{figure=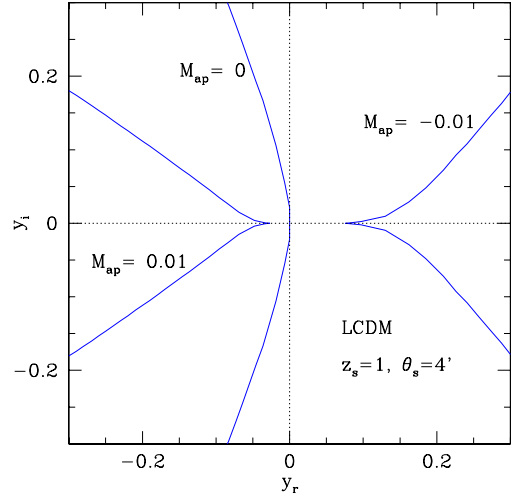,width=7cm,height=7cm}
\end{center}
\caption{The integration path over the complex plane for the aperture-mass
$\Map$ for the LCDM scenario, the source redshift $z_s=1$ and the angular
radius $\theta_s=4'$. We show the paths associated with $\Map=-0.01$ (right),
$\Map=0$ (center) and $\Map=0.01$ (left). We clearly see that for large
$|\Map|$ the integration path gets stucked onto one of the branch cuts of the 
generating function $\phiMap(y)$.}
\label{Figypath}
\end{figure}

In order to compute the pdf $\cP(\X)$ from the inverse Laplace transform
(\ref{PX}) we must perform the integration over $y$ in the complex plane.
We choose the integration path so that the argument of the exponential in
eq.(\ref{PX}) is a real negative number in order to avoid oscillations and
to obtain a fast convergence. From the definition (\ref{phiX}) it is clear
that the generating function $\phiX(y)$ obeys the symmetry 
$\phiX(y^*)=\phiX(y)^*$, for any real observable $\X$ like the shear component
$\gamma_{1s}$ or the aperture-mass $\Map$, so that the integration path 
over $y$
is symmetric with respect to the real axis. Moreover, we have to make sure
that the integration path does not cross the branch cuts of $\phiX(y)$. This
implies that for large $|\X|$ the integration path is pinched on the real axis
onto the singularities $y_{s,\X}^{\pm}$. This directly yields an exponential
tail for the pdf $\cP(\X)$, as can be seen in a straightforward way from the
expression (\ref{PX}). We discussed this point in more detail in 
Sect.~\ref{Exponential tails} above. For illustration, we show in 
Fig.~\ref{Figypath} the integration paths obtained for the aperture-mass
in the LCDM case for $z_s=1$ and $\theta_s=4'$. For $\Map=0$ the path runs
through the origin $y=0$ while for large positive $\Map$ it gets stucked onto
the negative branch cut of $\phiX(y)$ (and similarly for large negative 
$\Map$).

\subsection{Edgeworth expansion}
\label{Edgeworth expansion}

As is well known, a centered Gaussian distribution is fully defined by its 
second order moment $\lag\X^2\rag_c$. When the pdf $\cP(\X)$ deviates from
the Gaussian all higher-order cumulants $\lag\X^p\rag_c$ generically become
non-zero and the pdf now depends on this whole series. Thus, in
Sect.~\ref{Minimal tree-model} and Sect.~\ref{Stellar model} we had to
resum all these cumulants through the generating function $\phiX(y)$ defined
in eq.(\ref{phiX}) in order to determine the pdf $\cP(\X)$ as in 
eq.(\ref{PX}). However, when the deviations from the Gaussian are small one
can use the asymptotic Edgeworth expansion which describes the departures
from the Gaussian encoded by the few lowest order cumulants. Let us define
the parameters $S_p^{\X}$ associated with the variable $\X$ as in 
eq.(\ref{phiSp}):
\beq
\phiX(y) = \sum_{p=2}^{\infty} \frac{(-1)^{p-1}}{p!} \; S_p^{\X} \; y^p 
\hspace{0.3cm} \mbox{with} \hspace{0.3cm} 
S_p^{\X} = \frac{\lag\X^p\rag_c}{\lag\X^2\rag_c^{p-1}} .
\label{phiSpX}
\eeq
Here we assumed the random variable $\X$ to have zero mean ($\lag\X\rag=0$)
as is the case for weak lensing observables from eq.(\ref{Xx}) since 
$\lag\delta\rag=0$. Then, substituting the expansion (\ref{phiSpX}) into
eq.(\ref{PX}) and expanding the non-Gaussian part of the exponent one
obtains (e.g., Bernardeau \& Kofman 1995):
\beqa
\cP(\X) & = & \frac{1}{\sqrt{2\pi\sigma_{\X}^2}} \; e^{-\nu^2/2} \; 
\Bigg \lbrace 1 + \sigma_{\X} \frac{S_3^{\X}}{6} H_3(\nu) \nonumber \\
& & + \sigma_{\X}^2 \left[ \frac{S_4^{\X}}{24} H_4(\nu) + 
\frac{(S_3^{\X})^2}{72} H_6(\nu) \right] + .. \Bigg \rbrace
\label{Edg}
\eeqa
with:
\beq
\sigma_{\X} = \lag\X^2\rag_c^{1/2} \hspace{0.3cm} \mbox{and} \hspace{0.3cm}
\nu = \frac{\X}{\sigma_{\X}} .
\label{nu}
\eeq
Here we introduced the Hermite polynomials $H_n(\nu)$. In particular we have:
\beq
H_3(\nu) = \nu^3 - 3 \nu \hspace{0.3cm} \mbox{and} \hspace{0.3cm}
H_4(\nu) = \nu^4 - 6 \nu^2 + 3 .
\label{Hnu}
\eeq
Since this is an asymptotic expansion, the Edgeworth expansion 
(\ref{Edg}) is only useful for moderate deviations from the Gaussian, that
is when the first correcting term is smaller than unity (typically $|\nu| 
\la 1$ and $|\sigma_{\X}S_3^{\X}|\la 1$), and the accuracy does not improve 
by including higher order terms in the expansion (\ref{Edg}) (unless the pdf is
extremely close to the Gaussian). The advantage of the Edgeworth 
expansion is that it provides a straightforward estimate of the pdf 
$\cP(\X)$ when it is still relatively close to the Gaussian, using only 
the skewness $S_3^{\X}$ or the kurtosis $S_4^{\X}$. This avoids the need 
to resum all higher-order cumulants and to perform the integration over 
the complex plane as in eq.(\ref{PX}).

\section{The shear and aperture-mass statistics from numerical simulations}
\label{Numerical simulations}

The numerical method for the computation of the lensing statistics is
based on the original formalism of Couchman, Barber \& Thomas, 1999,
and which has been further developed by Barber, 2002. The original
formalism allows for the computation of the three-dimensional shear
matrices at a large number of locations within each $N$-body
simulation volume output. In the present work we have evaluated the
shear at 300 locations along every one of $455 \times 455$ lines of sight in
each of the simulation volumes. The development allows for the
successive combination of these matrices along the lines of sight and
throughout the linked simulation volumes from the sources at the
required redshifts to the observer at $z =0$. The overall procedure
therefore gives rise to Jacobian matrices for each line of sight and
for each of the specified source redshifts from which the required
lensing statistics have been generated.

Our procedure has been applied to two different cosmological
simulations created by the Hydra
Consortium\footnote{(http://hydra.mcmaster.ca/hydra/index.html)} using
the `Hydra' $N$-body hydrodynamics code (Couchman, Thomas \& Pearce,
1995). The parameters describing the two cosmologies, LCDM and OCDM,
are given in Table 1. Both contained $86^3$ dark matter particles of
mass $1.29 \times 10^{11}h^{-1}$ solar masses each, where $h$ is the
value of the Hubble parameter expressed in units of
100~km~s$^{-1}$~Mpc$^{-1}$. We used a variable particle softening
within the code, to reflect the density environment of each particle,
the minimum value of which, for particles in the densest environments,
was chosen to be $0.0007(1+z)$ in box units, where $z$ is the redshift
of the particular simulation volume. 
\begin{table}
\begin{center}
\caption{The parameters used in the two cosmological
simulations. $\Om$ is the matter density parameter, $\Ol$ is the
vacuum energy density parameter, $\Gamma$ is the power spectrum shape
parameter, $\sigma_8$ is the normalisation on scales of $8h^{-1}$~Mpc,
$\theta_{\rm res}$ is the angular resolution and $\theta_{\rm
survey}^2$ is the angular size of the complete field of view
throughout the simulations.}
\label{tabsig2D}
\begin{tabular}{@{}lcccccc}
\hline
&$\Om$&$\Ol$&$\Gamma$&$\sigma_8$&$\theta_{\rm res}$&$\theta_{\rm survey}^2$\\
\hline
LCDM&0.3&0.7&0.25&1.22&$0'.34$&$2^{\circ}.6 \times 2^{\circ}.6$ \\
OCDM&0.3&0.0&0.25&1.06&$0'.37$&$2^{\circ}.8 \times 2^{\circ}.8$ \\
\hline
\end{tabular}
\end{center}
\end{table}

The angular resolution in the LCDM cosmology was $0'.34$, equivalent
to the minimum value of the particle softening at $z = 0.36$, which is
the redshift for the maximum lensing effects for sources at a redshift
of 1 in that cosmology. In the case of the OCDM cosmology, the angular 
resolution was $0'.37$. The angular size of the survey
referred to in Table 1 corresponds to completely filling the front
face of the redshift 1 simulation volume for an observer at redshift
zero. For source redshifts greater than 1, the periodicity of the
particle distributions was used to allow lines of sight beyond the
confines of the simulation volumes to be included, as described in
Barber, Munshi \& Valageas, 2003.

The sources for our simulations were assumed to lie in a regular
$455 \times 455$ array at the front face of the simulation volume corresponding
to the specified redshift. We used 14 different values for the source redshifts in
each cosmology, whose exact redshift values are specified in Table 2.
The redshifts selected were chosen to be close to redshifts of 0.1,
0.2, 0.3, 0.4, 0.5, 0.6, 0.7, 0.8, 0.9, 1.0, 1.5, 2.0, 3.0 and 3.5. In
this paper the redshifts are referred to loosely as these approximate
values, although in the determination of the lensing statistics, the
actual redshift values were used.

\begin{table*}
\begin{center}
\caption{The redshifts of the sources in the two cosmologies.}
\label{tabsig2D}
\begin{tabular}{@{}lcccccccccccccc}
\hline
&$z_1$&$z_2$&$z_3$&$z_4$&$z_5$&$z_6$&$z_7$&$z_8$&$z_9$&$z_{10}$&$z_{11}$&$z_{12}$&$z_{13}$&$z_{14}$\\
\hline
LCDM&0.10&0.21&0.29&0.41&$0.49$&$0.58$&0.72&0.82&0.88&.99&1.53&1.97&3.07&3.57 \\
OCDM&0.11&0.18&0.31&0.41&$0.51$&$0.63$&0.69& - &0.88&1.03&1.47&2.03&3.13&3.53 \\
\hline
\end{tabular}
\end{center}
\end{table*}

Each of the simulation volumes had comoving side-dimensions of
100$h^{-1}$Mpc and to avoid obvious structure correlations, each was
arbitrarily translated, rotated and reflected about each coordinate
axis for each of the total of $N=10$ runs through all the volumes. In
this way, we obtained $N$ sets of Jacobian matrices for the different
runs. 

\begin{figure}
\protect\centerline{
\epsfysize = 2.truein
\epsfbox[5 5 250 250]
{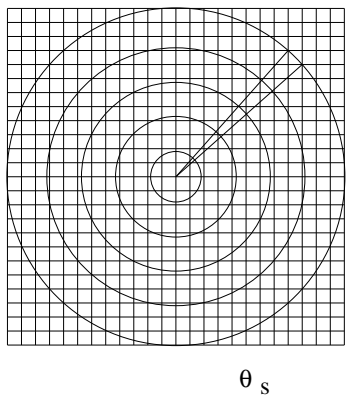}}
\caption{Schematic description of the angular bins used for evaluation of
$M_{ap}$ statistics.}
\label{filter.eps}
\end{figure}

By extracting the data needed from the Jacobians, the data for the
shear components and convergence were smoothed on the different
angular scales using a top-hat filter for the shear and the
compensated filter for the aperture mass. To obtain the required pdfs
and lower order moments the unsmoothed field in real space was
convolved with the appropriate filters using the following
scheme. Concentric annuli, separated equally in the radial direction,
were used for sampling of the denisty field, as shown in
figure~\ref{filter.eps}. These annuli were than divided into equally
spaced angular bins. The density field was computed along the radial
and angular bins using a linear interpolation scheme from adjacent
bins, before convolving it with the filters. Various combinations of
the radial bin width and angular bin size were considered to check the
convergence and numerical stability of the scheme. Our scheme enforces
the radial symmetry inherent in the window functions which is
important when dealing with window functions that have more features
compared to a tophat filter. We have checked the stability by taking
up to $20$ radial binning divisions for large smoothing radius and
$60$ divisons along the angular directions. In addition, we have given
random shifts to the rectangular grid along both axes to get a better
sampling of the density field. We have also checked various levels of
dilution by considering randomly selected points in the catalogue. In
addition, we have checked various levels of uniform dilution where we
choose uniformly spaced points to evaluate the statistics of the
aperture mass, $\Map$. It became clear from our studies that grid
effects will be much more pronounced in any statistical analysis
related to compensated filters. This is due to the fact that for a
given smoothing scale, compensated filters take more contributions
from smaller angular scales.

Finally, the computed values for the pdfs, higher-order moments and
aperture mass from each of the runs in each cosmology were averaged so
that the errors on the means of $1\sigma/\sqrt{N}$ for each statistic
were determined.

\section{Results}
\label{Results}

Our study can be divided in two parts.
In addition to the lower order moments we compute the
complete pdf and we study their variation as a function of smoothing radius
as well as their evolution with redshift. For the pdf of the aperture mass
we investigate both the minimal tree model and the stellar model, which 
allows us to study the dependence on the detailed modeling of the 
correlation function (both models give close results for the shear).

\subsection{The length scales probed by shear statistics}

\begin{figure}
\psfig{figure=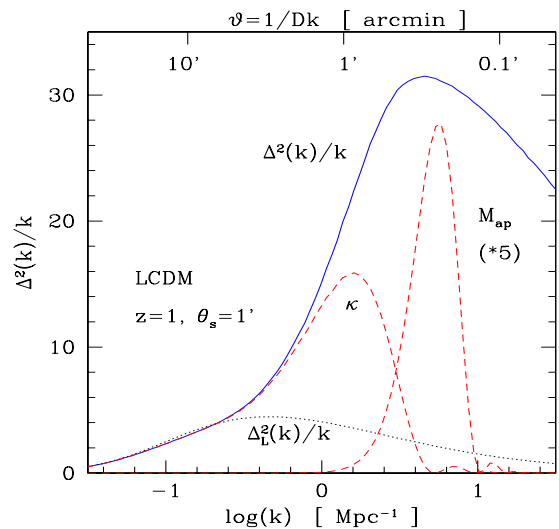,width=8cm,height=7cm}
\caption{The contribution of various comoving wavenumbers $k$ to weak lensing
observables at redshift $z=1$ along the line of sight. The solid line
is the non-linear power $\Delta^2(k)/k$ (obtained from Peacock \& Dodds 1996)
while the dotted line is the linear power $\Delta^2_L(k)/k$ (i.e. using the
linear power-spectrum). The left dashed line shows the contribution 
$\Wkappa(k_{\perp}\De\theta_s)\Delta^2(k)/k$ to the variance of the smoothed 
convergence $\kappa_s$ (or to the smoothed shear component $\gamma_{1s}$)
for the angular radius $\theta_s=1'$. The right dashed line shows the 
contribution $\WMap(k_{\perp}\De\theta_s)\Delta^2(k)/k$, multiplied by a 
factor $5$, to the variance of the aperture-mass $\Map$ 
for $\theta_s=1'$. The upper axis shows the angular scale $\vartheta=1/\De k$
associated to comoving wavenumber $k$.}
\label{FigD2k}
\end{figure}

As we study a range of source redshifts and smoothing angular scales
the weak lensing observables (such as the shear or the aperture-mass)
probe the density field over various length scales and redshifts, which
run from the linear to the highly non-linear regime.
The redshift dependence of the typical length scale comes from the angular 
diameter distance ${\cal D}(z)$, see eq.(\ref{ks}). 
The typical comoving wavenumber probed
by various shear statistics for a smoothing angle $\theta_s=1'$ are of the
order of $1$ to $10h {\rm Mpc}^{-1}$ which corresponds to scales
$0.1$ to $1 h^{-1}$ Mpc, see Fig.5 in Barber, Munshi \& Valageas (2003). In terms of
the power per logarithmic wavenumber interval $\Delta^2(k)$, for 
$1'<\theta_s<8'$, this yields $0.1 \la \Delta^2(k) \la 400$, see
Fig.6 in Barber, Munshi \& Valageas (2003). Hence most of the contribution to the
weak lensing shear comes from intermediate length scales which cannot
be described by the quasi-linear limit nor by the stable-clustering ansatz.
As a consequence, in order to study weak lensing over the angular scales of
interest for observational purposes (i.e. from $1'$ up to $20'$) one needs to
use a model which can be applied from the quasi-linear regime up to the
highly non-linear regime. The models which we recalled in 
Sect.~\ref{The density field} (see also Valageas, Barber \& Munshi 2003 and 
Barber, Munshi \& Valageas 2003 for details) provide such a tool and yield all 
properties of the density field (both the amplitude and the angular dependence
of the many-body correlations) over the entire dynamical range.

We show in Fig.~\ref{FigD2k} the contribution of various comoving wavenumbers 
$k$ to weak lensing observables at redshift $z=1$ along the line of sight.
We first plot the non-linear power $\Delta^2(k)/k$ (obtained from Peacock 
\& Dodds 1996, solid line) and the linear power $\Delta^2_L(k)/k$ (i.e. using 
the linear power-spectrum, dotted line). Indeed, because of the projection
associated with the integration along the line-of-sight, the power coming from
a wavenumber $k$ to weak lensing observables is not $\Delta^2(k)$ (as for the
3-d density contrast $\delta_R$) but $\Delta^2(k)/k$, as seen from 
eq.(\ref{varX}). For CDM-like power-spectra, which flatten at small scales,
we can see that the power $\Delta^2(k)/k$ decreases beyond $6$ Mpc$^{-1}$.
Next, we show the contribution $\Wkappa(k_{\perp}\De\theta_s)\Delta^2(k)/k$ 
to the variance of the smoothed convergence $\kappa_s$ (or to the smoothed 
shear component $\gamma_{1s}$) for the angular radius $\theta_s=1'$ (left
dashed line). At small wavenumbers it follows the power $\Delta^2(k)/k$ 
since $\Wkappa(0)=1$ and it exhibits a cutoff beyond $1/(\De\theta_s)$.
The right dashed line is the contribution 
$\WMap(k_{\perp}\De\theta_s)\Delta^2(k)/k$ to the variance of the 
aperture-mass $\Map$ for the same angular radius $\theta_s=1'$,
multiplied by a factor 5 (for clarity on the figure).
We can see that it probes a much narrower range of wavenumbers since the
contribution from long wavelengths is suppressed because it involves
a compensated filter with $\WMap(0)=0$. Moreover, for the same angular radius
we can check that $\Map$ probes higher wavenumbers than $\kappa_s$
because of this suppression of long wavelengths and of the profile of the
filter which varies over scales $\sim \theta_s/4$. More precisely,
from the upper axis which shows the angular scale $\vartheta=1/\De k$ 
associated with comoving wavenumber $k$, we see
that the smoothed convergence or the smoothed shear mainly probe the density
field at the wavenumber $k_s=1/\De\theta_s$ while contributions to the
aperture-mass peak around $k_s=4/\De\theta_s$. This is why we chose
the normalization (\ref{ks}) for the typical wavenumber $k_s$.
Finally, the curves in Fig.~\ref{FigD2k} clearly show that the weak lensing
signal comes from non-linear scales which cannot be described by the linear
power $\Delta^2_L(k)/k$.

\begin{figure}
\protect\centerline{
\epsfysize = 1.8truein
\epsfbox[18 400 588 715]
{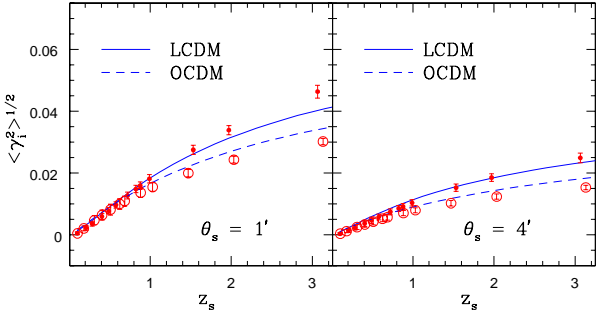}}
\caption{Variance of the smoothed shear component $\gamma_{is}$, 
$\lag \gamma_{is}^2 \rag^{1/2}$, as a function of the source redshift $z_s$. 
The smoothing angle $\theta_s$ is fixed at $1$ arcminute in left panel and 
at $4$ arcminute in the right panel. Lines correspond to the analytical 
prediction (\ref{varX}) and data points represent results from numerical 
simulations. Error bars are computed from scatter among various realizations.
Variance of shear increases with source redshift $z_s$ for a fixed smoothing 
angle. Dots represent LCDM simulations whereas
OCDM simulations are represented by open circles.}
\label{var_red}
\end{figure}

\begin{figure}
\protect\centerline{
\epsfysize = 1.8truein
\epsfbox[18 400 588 715]
{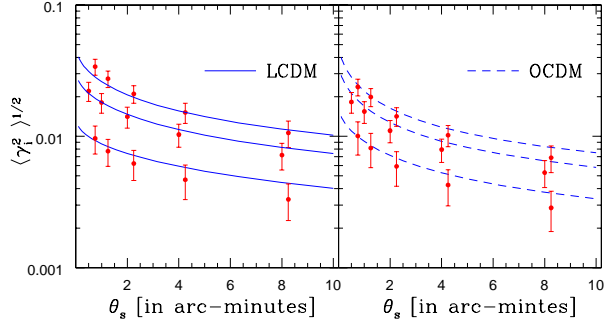}}
\caption{Variance is plotted as a function of smoothing angle for various fixed
redshifts. Left panel correspond to LCDM and the right panel correspond to 
OCDM cosmology. The solid (LCDM) and dashed (OCDM) lines in each panels denote 
the analytical prediction (\ref{varX}) for the variance at redshifts
$z_s = 0.5, 1$ and $1.5$ (from bottom to top). Points with error bars are 
numerical results from simulations.}
\label{var_theta}
\end{figure}

\begin{figure}
\protect\centerline{
\epsfysize = 1.8truein
\epsfbox[18 400 588 715]
{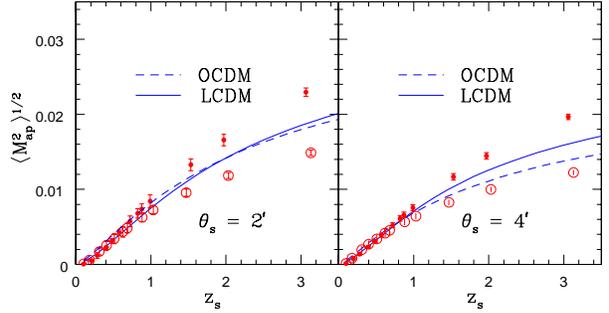}}
\caption{Variance plotted as a function of source redshift $z_s$.
The aperture-mass compensated filter ($\Map$) is used to smooth the shear 
map. Left panel corresponds to
smoothing angle $\theta_s = 2'$ and the right panel corresponds to 
$\theta_s = 4'$. The solid lines 
correspond to LCDM cosmology whereas the dashed lines correspond to
OCDM cosmology. Data points represent averages of $10$ different 
realizations for each cosmology. Error bars are computed from scatter 
among various realizations. Dots represent LCDM simulations whereas
OCDM simulations are represented by open circles.}
\label{map_var_red}
\end{figure}

\begin{figure}
\protect\centerline{
\epsfysize = 1.8truein
\epsfbox[18 400 588 715]
{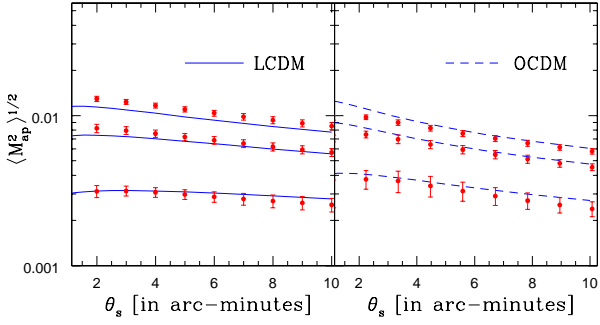}}
\caption{Aperture-mass variance plotted as a function of smoothing angle 
$\theta_s$. Left panel corresponds to
LCDM cosmology whereas the right panel corresponds to OCDM cosmology. Three
different redshifts are considered: $z_s = 1.5, 1$ and $z_s=0.5$ from top 
to bottom. Data points
represent averages of $10$ different realizations for each cosmology.
Error bars are computed from scatter among various realizations. }
\label{map_var_theta}
\end{figure}

\subsection{Variance of Shear and Aperture-Mass}
\label{Variance of Shear and Aperture-Mass}

We study the variance of smoothed shear components 
$\lag \gamma_{is}^2\rag^{1/2}$ and aperture-mass $\lag \Map^2 \rag^{1/2}$
as a function of both source redshift $z_s$ and smoothing angle $\theta_s$
in Figs.~\ref{var_red}-\ref{map_var_theta}.
It follows the increase with redshift of the length of the line of sight. 
The variance is smaller in case of OCDM model
mainly because of the smaller normalization $\sigma_8$. 
This implies that the LCDM pdfs are broader compared to
OCDM pdfs for a given smoothing angle and a given redshift.
As in the case of our convergence studies we have a reasonable agreement
from the fitting formula of Peacock \& Dodds (1996) over the entire
range of redshift and smoothing angle.
%
%
Our results agree with previous 
studies which focused on the variance of compensated filters 
(e.g., Hoekstra et al. (2002), van Waerbeke et al. (2002), 
Jarvis et al. (2003), Hamana et al. (2003), Benabed \& van Waerbeke (2003)).
However, we can see that at large redshifts and small scales, there appears
to be some discrepancy between the analytic results and the simulations.
This deviation shows the same behaviour for the shear and the aperture-mass
but it is larger for the latter. We must point out that the computation of 
variance only depends on the two-point density correlation function 
(or the power spectrum) and not on the entire correlation hierarchy or the 
parameters $S_p$, which are the main focus of this paper. In other words,
the deviations seen at large $z_s$ and small $\theta_s$ in 
Fig.~\ref{map_var_red} are solely due to the mismatch between the fit from
Peacock \& Dodds (1996) to the power spectrum and the numerical simulation.
Therefore, they probably signal the effects of the finite numerical resolution
(note that the softening length increases with $z$). On the other hand, it
is known that the fit from Peacock \& Dodds (1996) is not perfect
(Smith et al. 2002).
%
%

\begin{figure}
\protect\centerline{
\epsfysize = 1.8truein
\epsfbox[18 400 588 715]
{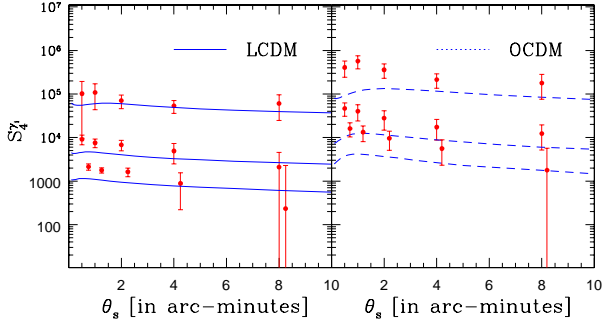}}
\caption{Kurtosis (see eq.(\ref{phiSpX}) for definition) of the smoothed shear component $\gamma_{is}$, 
 as a function of the smoothing angle $\theta_s$.
The left panel corresponds to LCDM cosmology (solid lines) and the
right panel to OCDM cosmology (dashed lines). Lines from top to bottom 
correspond to analytical predictions (\ref{cumXk2}) (stellar model) 
for different source redshifts: $z_s = 0.5, 1$ and $1.5$. Points with 
error bars are the numerical results from simulations.}
\label{kurt_theta}
\end{figure}

\begin{figure}
\protect\centerline{
\epsfysize = 1.8truein
\epsfbox[18 400 588 715]
{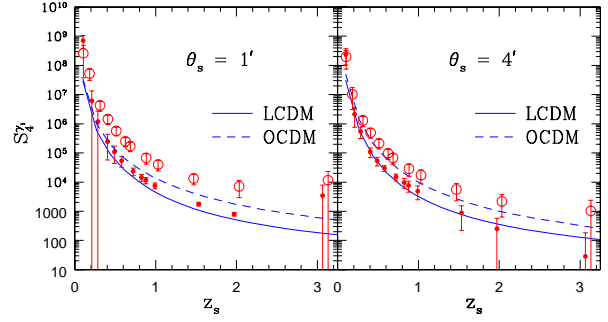}}
\caption{Kurtosis (see eq.(\ref{phiSpX}) for definition) of the smoothed shear component $\gamma_{is}$
as a function of the source redshift $z_s$. 
The smoothing angle $\theta_s$ is fixed at $1$ arcminute in the left panel and 
at $4$ arcminute in the right panel. Lines correspond to the analytical 
prediction (\ref{cumXk2}) (stellar model) and data points represent results 
from numerical simulations. Error bars are computed from scatter among 
various realizations. Although shear variance increases with redshift the 
kurtosis decreases with source redshift $z_s$. Black dots with error bars 
represent LCDM simulation while open circles represent OCDM simulations. }
\label{kurtz}
\end{figure}

\subsection{The Kurtosis of Shear}
\label{The Kurtosis of Shear}

As pointed out in Sect.~\ref{Weak lensing distortions}, for symmetry reasons 
all odd-order cumulants of shear components vanish. Therefore, the 
lowest-order non-Gaussian contribution is the kurtosis which we have studied 
as a function of both smoothing angle $\theta_s$ (Fig.~\ref{kurt_theta}) and
source redshift $z_s$ (Fig.~\ref{kurtz}) (see also Takada \& Jain 2002).
%
%
We define the kurtosis as $S_4^{\gamma_{is}}=\lag\gamma_{is}^4\rag_c/
\lag\gamma_{is}^2\rag^3$, as in eq.(\ref{phiSpX}).
%
%
We can check that we obtain a reasonable agreement with the numerical 
simulations. There is some discrepancy at small scales but this might be 
related to the numerical resolution. On the other hand, it is increasingly
difficult to predict with a good accuracy higher-order moments of weak-lensing
observables (or of the density field itself). In particular, while our model
for the density field is in a sense ``fitted'' to the skewness $S_3$ of the
density contrast through (\ref{S3}) (the quasi-linear limit $S_3^{\rm QL}$
is exact while the non-linear HEPT ansatz $S_3^{\rm NL}$ was seen to agree
with numerical simulations), this is not the case for the kurtosis and 
higher-order moments of the density contrast. They are set by the simple 
parameterization of the generating function $\varphi(y)$ described in
Valageas, Barber \& Munshi (2003) and Barber, Munshi \& Valageas (2003) which only depends on $S_3$.
Moreover, in the case of weak-lensing observables there is a further dependence
on the angular behaviour of the correlation functions. Thus, the minimal
tree-model and the stellar model do not give identical results for the 
kurtosis (since the former writes the four-point correlation as a sum of 
a ``stellar'' and a ``snake'' diagram while the latter only keeps the 
``stellar'' graph). Nevertheless, for the shear kurtosis both results are 
close and we only plot in Fig.~\ref{kurt_theta} and Fig.~\ref{kurtz} the 
stellar model prediction since it is this model which we shall investigate 
in more details in Sect.~\ref{Probability Distribution Function for Shear} 
for the full pdf because it is much more convenient for numerical purposes. 

Note that whereas the variance of the shear components increases with the 
source redshift the kurtosis decreases. This is due both to the longer 
length of the line of sight (as we add the lensing contributions from 
successive mass sheets along the line of sight the total signal becomes closer
to Gaussian, in agreement with the central limit theorem) and to the fact that
the density field is closer to Gaussian at higher redshift. We can see that
the variation with redshift of the kurtosis is actually very steep. This
implies that any realistic observational study which aims to determine 
cosmological parameters from the lower-order moments of the shear components
will have to determine source redshifts very accurately. On the other hand,
the analytical calculations must take into account the spread of the
distribution of source redshifts. This is actually straightforward within
the formalism used in this article (see Valageas (2000a)) if we neglect
non-linear couplings.

In realistic surveys one must also take into account the finite size of the 
catalog as well as the observational noise. A complete error analysis of 
higher-order moments for aperture-mass has been developed in Munshi \& Coles 
(2002). Clearly from observational
view points it is important to have a reasonable sky coverage and
dense sampling of galaxies. Note that error analysis of lower order moments 
of shear components involves incorporating correlations among neighbouring
cells which can be ignored for the study of aperture-mass statistics.
This makes such studies for shear components more complicated although
a crude order of magnitude study can be performed. It is also important to 
note that throughout we have ignored the
noise due to intrinsic ellipticity distribution which we have to
include in a more realistic study.

On the other hand, with future weak lensing surveys such as LSST and SNAP 
it will be feasible to study the non-Gaussianity induced by gravity with an
unprecedented accuracy. It was pointed out by Hui (1999) that the skewness 
of the convergence field can directly be used to probe the dark energy 
equation of state. Future surveys with large sky coverage and dense 
sampling can also use the kurtosis of shear components to study the 
properties of dark energy. A clear detection of kurtosis will also help us 
to break the degeneracies in determining $\Om$ and $\sigma_8$ which are 
inherent in studies based on power spectrum analysis only (see also 
Takada \& Jain 2002).

\begin{figure}
\protect\centerline{
\epsfysize = 1.8truein
\epsfbox[18 400 588 715]
{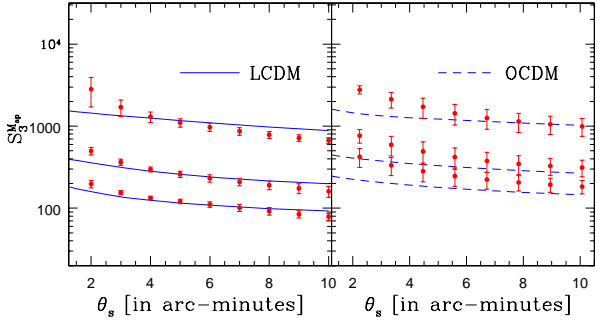}}
\caption{Skewness (see eq.(\ref{phiSpX}) for definition) of the 
aperture-mass $\Map$ plotted as a function of 
smoothing angle $\theta_s$. Lines show the analytical prediction for
$z_s = 0.5, 1$ and $z_s=1.5$ from top to bottom (the tree-model (\ref{cumX}) 
and the stellar model (\ref{cumXk2}) are identical for the skewness).
Left panel corresponds to LCDM cosmology whereas the right panel corresponds 
to OCDM cosmology. Data points represent averages of $10$ different 
realizations for each cosmology. Error bars are computed from scatter 
among various realizations.}
\label{map_skew_theta}
\end{figure}

\begin{figure}
\protect\centerline{
\epsfysize = 1.8truein
\epsfbox[18 400 588 715]
{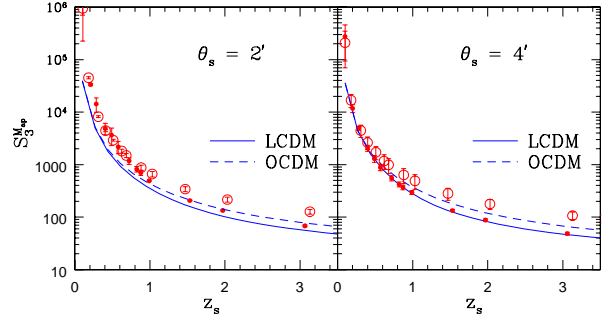}}
\caption{Skewness (see eq.(\ref{phiSpX}) for definition) of the 
aperture-mass $\Map$ plotted as a function of 
source redshift $z_s$. The left panel corresponds to
smoothing angle $\theta_s = 2'$ and the right panel corresponds to
$\theta_s = 4'$. The solid lines in each panel
correspond to LCDM cosmology whereas the dashed lines correspond to
OCDM cosmology. Data points represent averages of $10$ different 
realizations for each cosmology. Error bars are computed from scatter 
among various realizations. Black dots corresponds to LCDM simulations
whereas open circles correspond to OCDM simulations.}
\label{map_skew_red}
\end{figure}

\subsection{Skewness of Aperture-Mass}
\label{Skewness of Aperture-Mass}

The shear kurtosis can be directly measured from shear maps hence it is
a natural tool to study the departure from Gaussianity of the density
field. However, as we noticed in Sect.~\ref{The Kurtosis of Shear} it suffers
from several drawbacks. First, predictions for the kurtosis of the density
field are less robust than for the skewness, which is a lower-order moment.
Moreover, the relationship between the third-order moments of the density
constrast and weak-lensing observables is stronger because the dependence 
on the angular behaviour of the correlation functions is smaller. Thus, all
tree-models (including both the minimal tree-model and the stellar model
studied in more details in this paper) give the same predictions for any
statistics of order three (e.g., the skewness of the aperture-mass or 
any three-point correlation) since at this order there is only one tree graph,
whose weight is fully defined by $S_3$. This makes the skewness of the
aperture-mass $S_3^{\Map}$ the most useful probe of the non-Gaussianity
of the density field. An additional advantage is that the aperture-mass
provides a very localised probe of the density field in Fourier space, 
as seen in Fig.~\ref{FigD2k}. This makes the comparison between theory and
observations more robust and more precise as we can probe a narrow range
of scales, hence a well-defined regime of gravitational clustering.

Therefore, we plot in Fig.\ref{map_skew_theta} and Fig.~\ref{map_skew_red}
the skewness $S_3^{\Map}$ as a function of smoothing angle $\theta_s$
and source redshift $z_s$. We again obtain a reasonable agreement between
our theoretical predictions and the numerical simulations. Of course, like
the kurtosis $S_4^{\gamma_{is}}$ of the shear the skewness decreases at higher
redshifts. Note however that this variation is significantly shallower which
is a useful property for observational purposes (the error introduced by the
measure of the source redshifts will be smaller).

Although present generation surveys do not provide a clear cosmological
signal for non-zero skewness, future surveys such as Supernova 
Anisotropy Probe (SNAP) and the Large-aperture Synoptic Survey 
Telescope (LSST) will be very useful in this direction. On the other hand,
various generalizations of the skewness $S_3^{\Map}$ have been proposed
recently along with fast computational techniques which can reduce 
large volumes of data delivered by future weak lensing surveys.
These generalisations can also handle 
decomposition into cosmological modes and non-gravitational modes induced
by systematics (see e.g. Jarvis, Bernstein \& Jain 2003).

Our studies are complimentary to other works which have focused on 
constraining  or determining various halo model parameters from such 
observational studies. Our approach is significantly different as it
is based on the many-body correlations of the density field themselves, rather
than relying on a decomposition of the matter distribution over a population
of virialized halos.

\begin{figure}
\protect\centerline{
\epsfysize = 3.25truein
\epsfbox[25 150 588 715]
{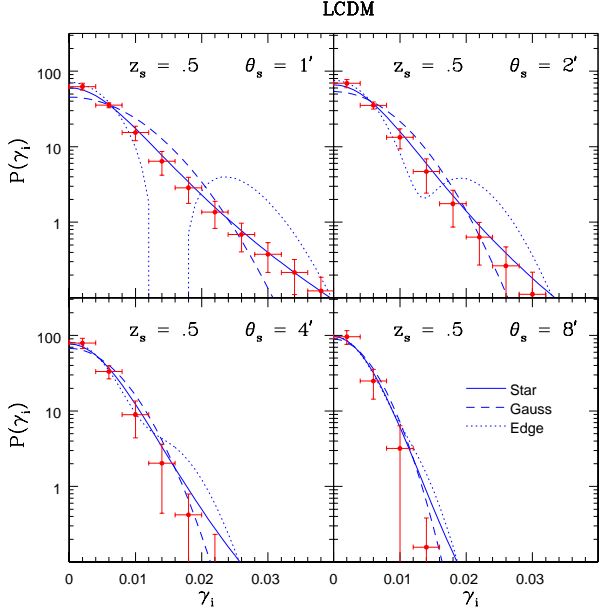}}
\caption{Probability distribution function $\cP(\gamma_{is})$ of the smoothed 
shear component $\gamma_{is}$. The smoothing angle $\theta_s$ is mentioned in 
each panel while the underlying cosmological parameters are that of LCDM. 
The sources are all placed at the same redshift $z_s = 0.5$. Since the pdf 
is even ($\cP(\gamma_{is})=\cP(-\gamma_{is})$) we only display the pdf 
for positive $\gamma_{is}$. Solid lines 
correspond to the analytical prediction (\ref{phiXstellar}) of the stellar 
model. The dashed lines show the Gaussian distribution with the same 
variance. The dotted lines correspond to the inclusion of the first non-zero 
term (kurtosis) in the Edgeworth expansion beyond the Gaussian approximation,
 eq.(\ref{Edg}). Data points are results from numerical simulations. Error 
bars indicate the scatter among various realizations.}
\label{lcdm_pdf_d5.eps}
\end{figure}

\begin{figure}
\protect\centerline{
\epsfysize = 3.25truein
\epsfbox[25 150 588 715]
{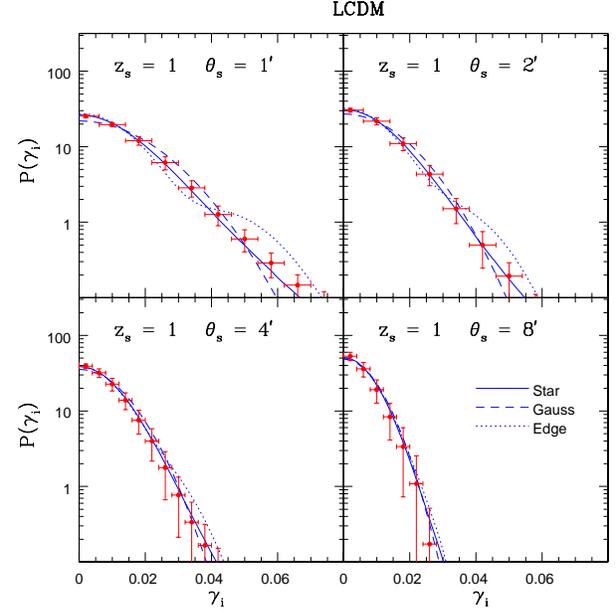}}
\caption{As in previous figure but for source redshift $z_s =1$. Note that
as we move towards higher redshifts the distribution becomes more Gaussian
and the Edgeworth expansion matches the analytical expression better. However,
truncating the expansion at any order produces spurious oscillations which
are not physical.}
\label{lcdm_pdf_1d.eps}
\end{figure}

\begin{figure}
\protect\centerline{
\epsfysize = 3.25truein
\epsfbox[25 150 588 715]
{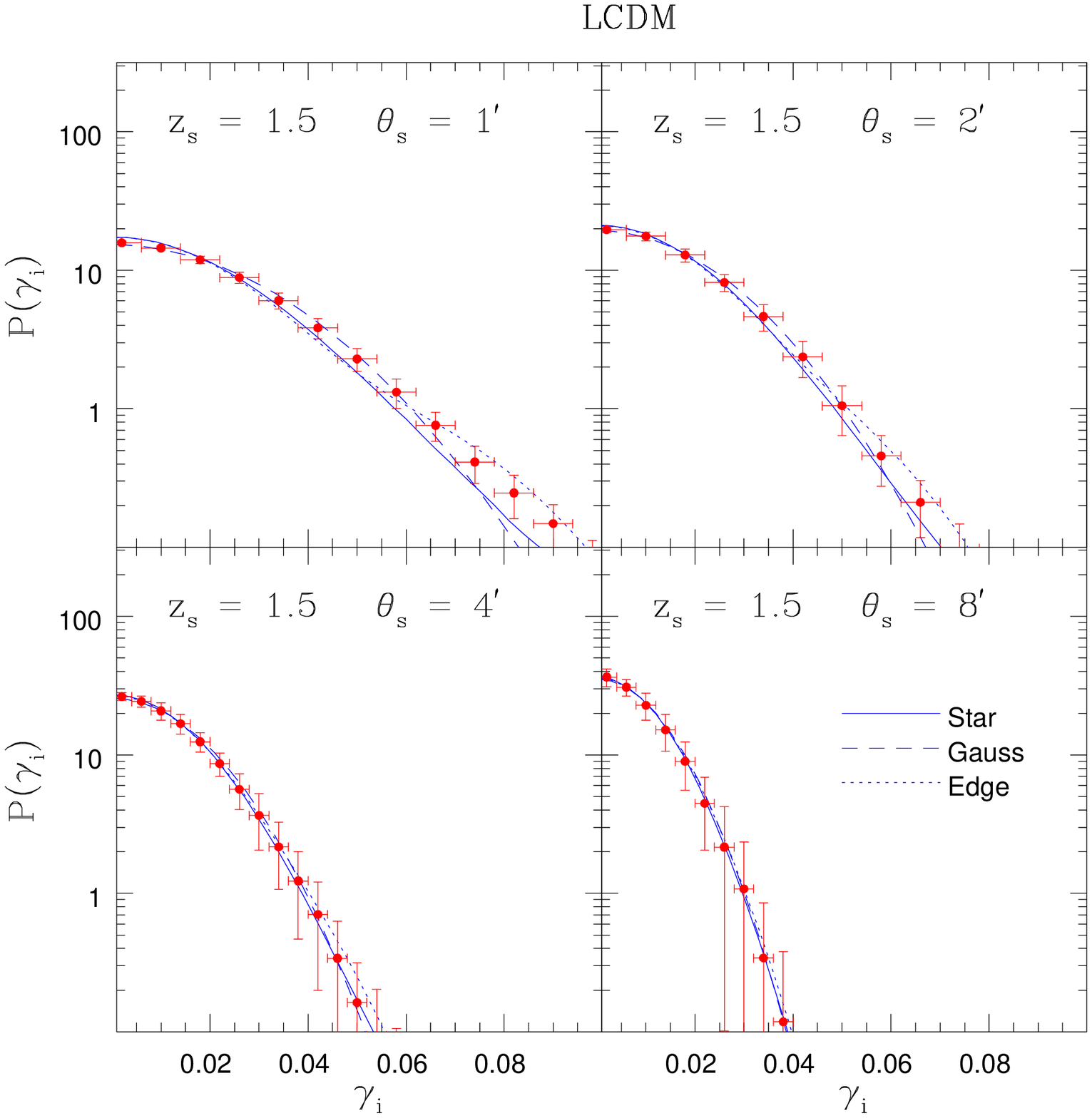}}
\caption{As in previous figure but for source redshift $z_s = 1.5$.}
\label{lcdm_pdf_1d5.eps}
\end{figure}

\begin{figure}
\protect\centerline{
\epsfysize = 3.25truein
\epsfbox[25 150 588 715]
{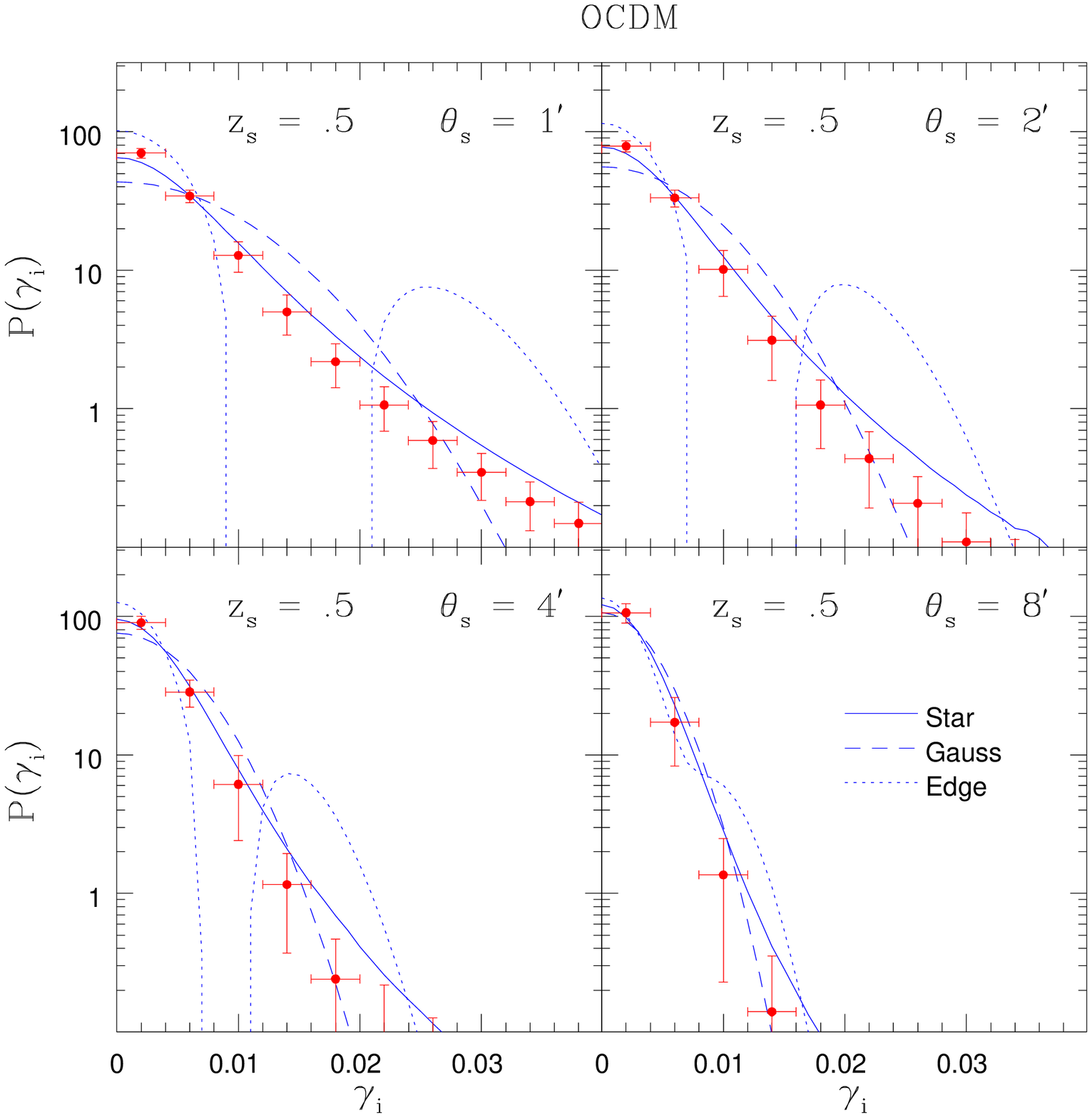}}
\caption{As in previous figure but for source redshift $z_s = 0.5$ and the
OCDM cosmology.}
\label{ocdm_pdf_d5.eps}
\end{figure}

\begin{figure}
\protect\centerline{
\epsfysize = 3.25truein
\epsfbox[25 150 588 715]
{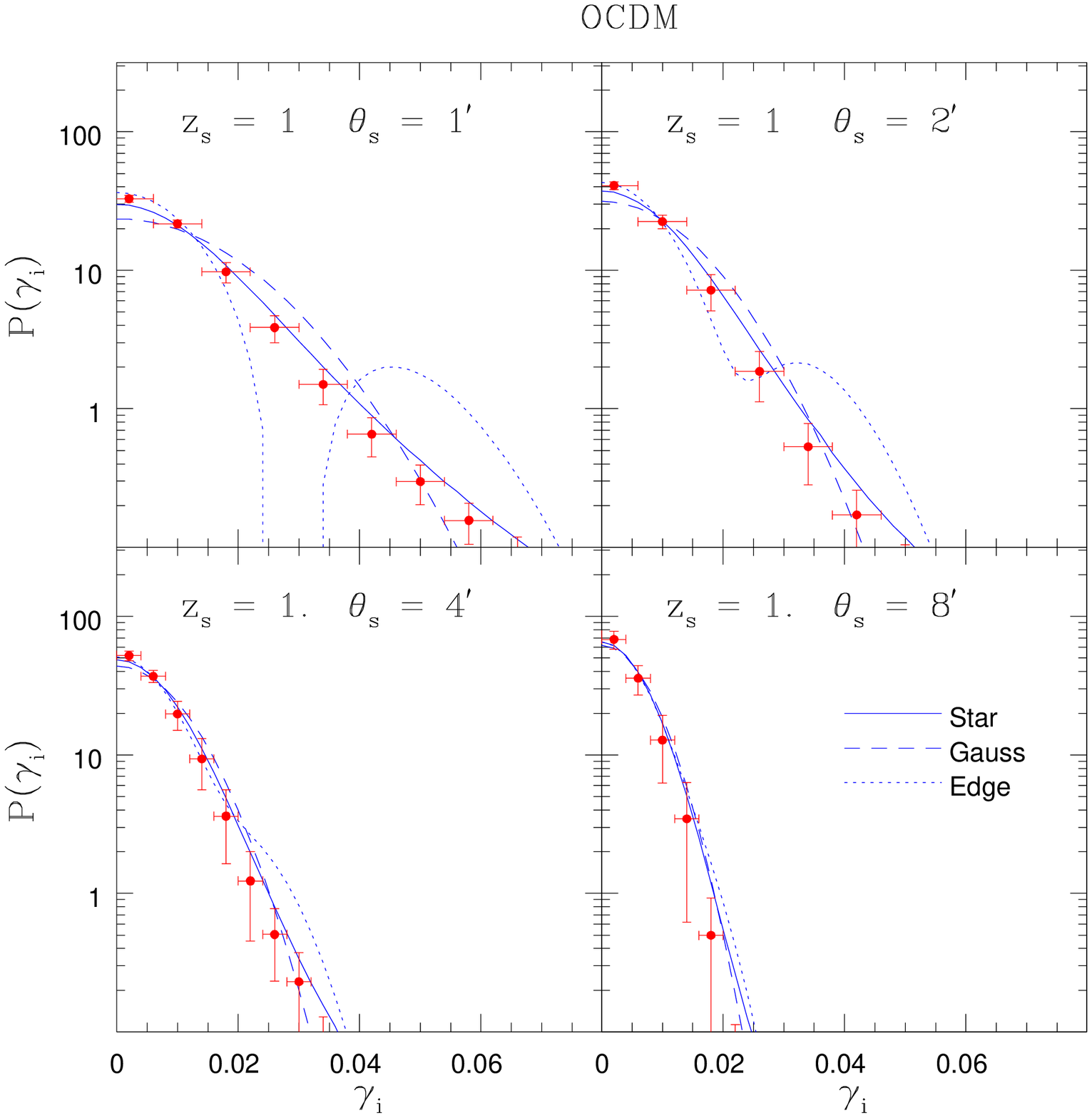}}
\caption{As in previous figure but for source redshift $z_s =1$.}
\label{ocdm_pdf_1d.eps}
\end{figure}

\begin{figure}
\protect\centerline{
\epsfysize = 3.25truein
\epsfbox[25 150 588 715]
{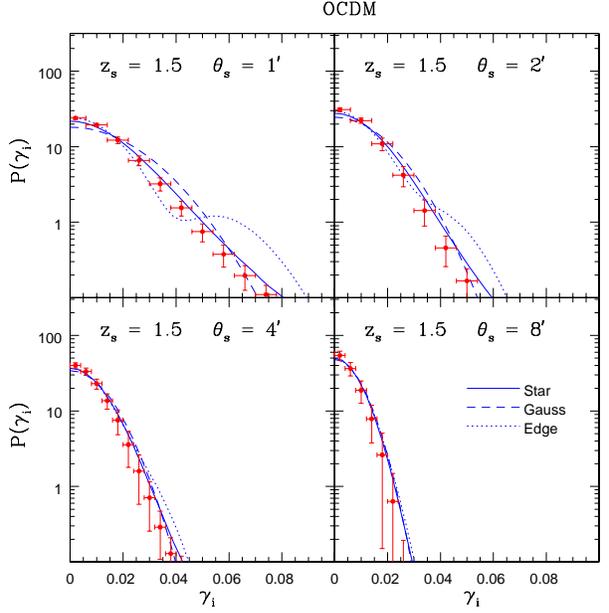}}
\caption{As in previous figure but for source redshift $z_s = 1.5$.}
\label{ocdm_pdf_1d5.eps}
\end{figure}

\subsection{Probability Distribution Function for Shear Components 
$\gamma_{is}$}
\label{Probability Distribution Function for Shear}

Our analytical results provide a complete analytical 
modeling of the pdf of shear components. Extending our earlier studies 
(Valageas, Barber \& Munshi 2003) we show that indeed such a 
description is possible for the entire range of redshifts as well as 
smoothing angles of practical interest. Thus, we compare with numerical
simulations our analytical predictions for the pdf $\cP(\gamma_{is})$ of the 
smoothed shear components in 
Figs.~\ref{lcdm_pdf_d5.eps} - \ref{ocdm_pdf_1d5.eps}.
We consider both LCDM and OCDM cosmologies, for smoothing angles 
$\theta_s = 1',2', 4'$ and $8'$ and for three different source redshifts
$z_s = 0.5, 1$ and $1.5$. We can see that we obtain a good agreement over
the entire range of smoothing radius and source redshift, from linear to
highly non-linear scales. We only plot the prediction of the stellar model
(\ref{phiXstellar}) because it is much more convenient for practical purposes.
Indeed, eq.(\ref{phiXstellar}) provides a simple explicit expression for the
generating function $\varphi_{\gamma_{is}}(y)$. By contrast, the minimal 
tree-model yields an implicit system (\ref{phiom})-(\ref{tauom}) which must
be solved numerically at each point $y$. Since the variable ${\vec \vartheta}$
is actually two-dimensional this involves solving for a 2-d function 
$\tau({\vec \vartheta})$ defined as the fixed point of eq.(\ref{tauom}). This
cannot be done by a simple iterative procedure for arbitary $y$ since it
would diverge for large $y$. Therefore, this implies the use of a 
time-consuming algorithm. Note that the generating function 
$\varphi_{\gamma_{is}}(y)$ must then be integrated over the complex plane in
eq.(\ref{PX}). Thus, the stellar model is much simpler to implement and since
both models give close predictions for the shear we focus on the stellar model
in Figs.~\ref{lcdm_pdf_d5.eps} - \ref{ocdm_pdf_1d5.eps}. Note that we have 
already shown in Barber, Munshi \& Valageas (2003) that both models also give almost
identical predictions for the smoothed convergence $\kappa_s$.

We also plot the Gaussian (dashed line) and the Edgeworth expansion (dotted
line) up to the first non-Gaussian correction (here the kurtosis) from
eq.(\ref{Edg}). We can check that at large smoothing angles $\theta_s$ 
and source redshifts $z_s$ the pdf $\cP(\gamma_{is})$ becomes very close to 
Gaussian. This could already be expected from the behaviour of the kurtosis
analysed in Sect.~\ref{The Kurtosis of Shear}. The departure from the 
Gaussian is significant at $z_s \sim 0.5$ and $\theta_s \la 2'$. On the other
hand, we note that the Edgeworth expansion is actually useless. It only
provides reasonable results when the pdf is very close to Gaussian while as
soon as there is a sizeable deviation from the Gaussian it introduces spurious
oscillations and it actually fares worse than the Gaussian. Note that since
it is an asymptotic expansion including higher-order terms would further
worsen this discrepancy.

One of the benefits of being able to study the shear components analytically
is to have a means to cross check the effects of various systematics which
go into the making of convergence maps from shear data. Alternatively one
can work with compensated filters to construct the $\Map$ statistics as
described above. Similarly, in a previous paper (Valageas, Barber \& Munshi 2003) we 
have shown that not only the statistics of the shear components can be 
modeled in this manner but one can also derive the properties of the
shear modulus $|\gamma_s|$. However, we found that the correlation between 
the two components is almost negligible so that the pdf of the shear
modulus does not carry useful additional information. This is the reason
why we have not included $\cP(|\gamma_s|)$ in the present study although
it is a simple outcome of the analytical results presented here.

Since the pdf $\cP(\gamma_{is})$ contains some information
about the entire hierarchy of the $S_p^{\gamma_{is}}$ cumulants it can be
used as a probe of the entire pdf of the underlying mass distribution.
This is clearly apparent in eq.(\ref{phiXstellar}). Moreover, it may be
used to distinguish cosmological models in a more efficient manner than
relying on the few lowest-order moments. Given that the shear maps are a 
direct outcome of any weak lensing survey, we may hope that surveys with 
a low level of noise as well as a good sky coverage could give us some clues 
on the distribution of matter in addition to the cosmological parameters.
However, we see in Figs.~\ref{lcdm_pdf_d5.eps} - \ref{ocdm_pdf_1d5.eps}
that the pdf $\cP(\gamma_{is})$ is not very far from the Gaussian, partly 
because
it is even so that the shape is roughly similar, hence it may be difficult to
extract accurate measures from future surveys.

\begin{figure}
\protect\centerline{
\epsfysize = 3.25truein
\epsfbox[25 150 588 715]
{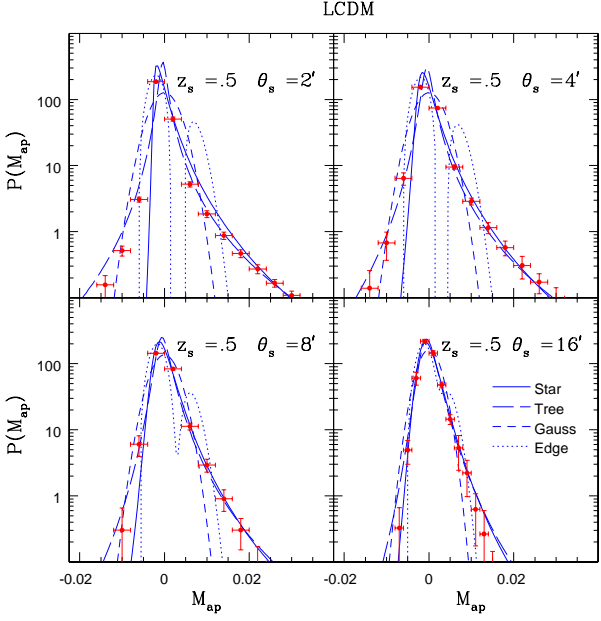}}
\caption{Probability distribution function $\cP(\Map)$ for the $\Map$ 
statistic which uses a compensated filter. We probe various smoothing radii
from $\theta_s =2'$ up to $\theta_s = 16'$. All sources are placed at source 
redshift $z_s = 0.5$ and we consider the LCDM cosmology. Solid lines show
the prediction of the stellar model (\ref{phiXstellar}). Long dashed lines
correspond to the minimal tree-model (\ref{phiom})-(\ref{tauom}). Short dashed
lines are the Gaussian with the same variance while the dotted lines are
the Edgeworth expansion (\ref{Edg}) including the first correction to 
Gaussianity (skewness). Data points are results from numerical simulations.
Ten different realizations are analysed and error bars denote the
scatter among various realizations. Note that for negative $\Map$ the 
minimal tree-model predictions are much closer to numerical simulations than
the stellar model.}
\label{pdf_map_lcdmzd5.eps}
\end{figure}

\begin{figure}
\protect\centerline{
\epsfysize = 3.25truein
\epsfbox[25 150 588 715]
{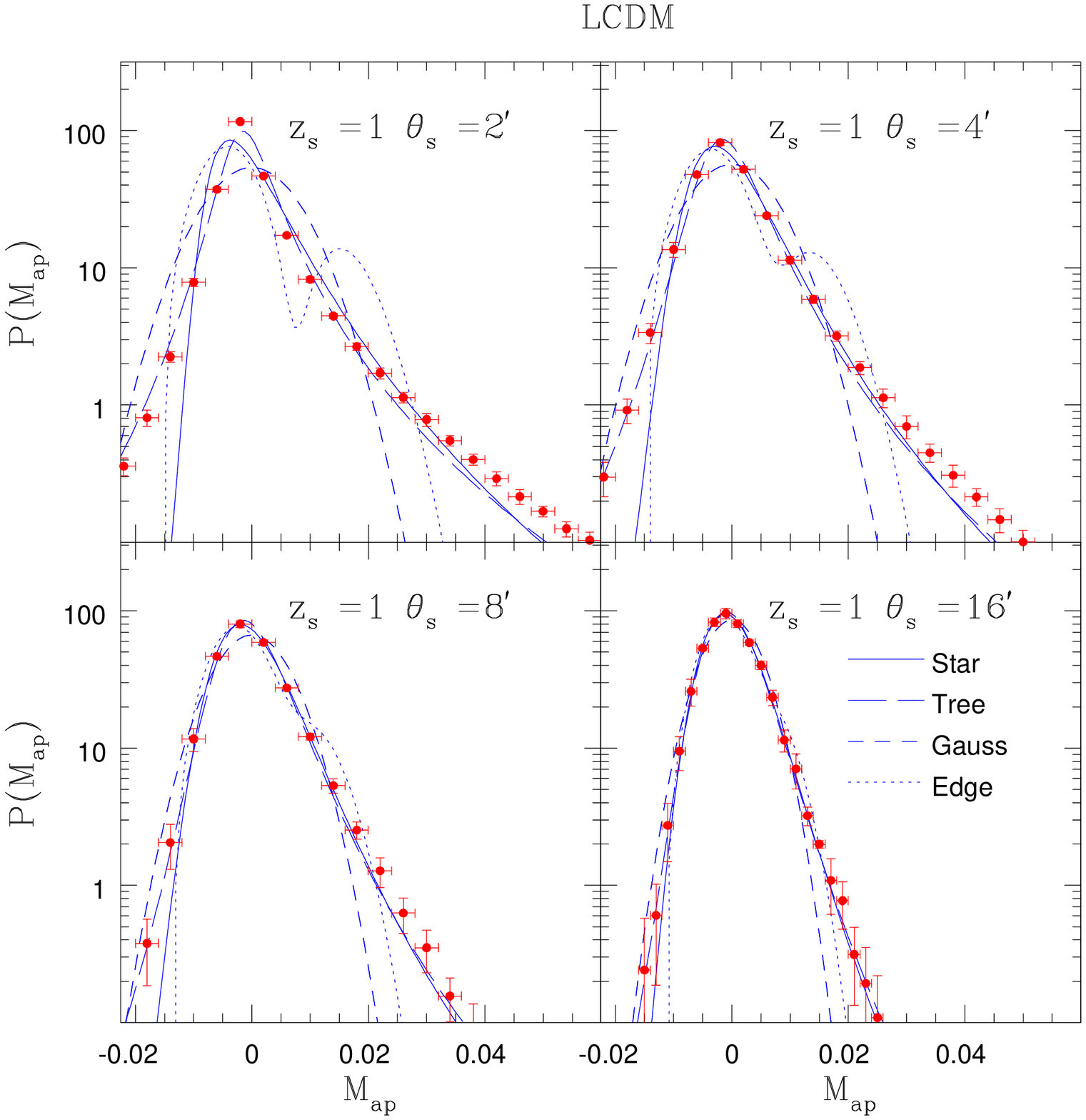}}
\caption{As in previous figure but for source redshift $z_s = 1$.}
\label{pdf_map_lcdmz1d.eps}
\end{figure}

\begin{figure}
\protect\centerline{
\epsfysize = 3.25truein
\epsfbox[25 150 588 715]
{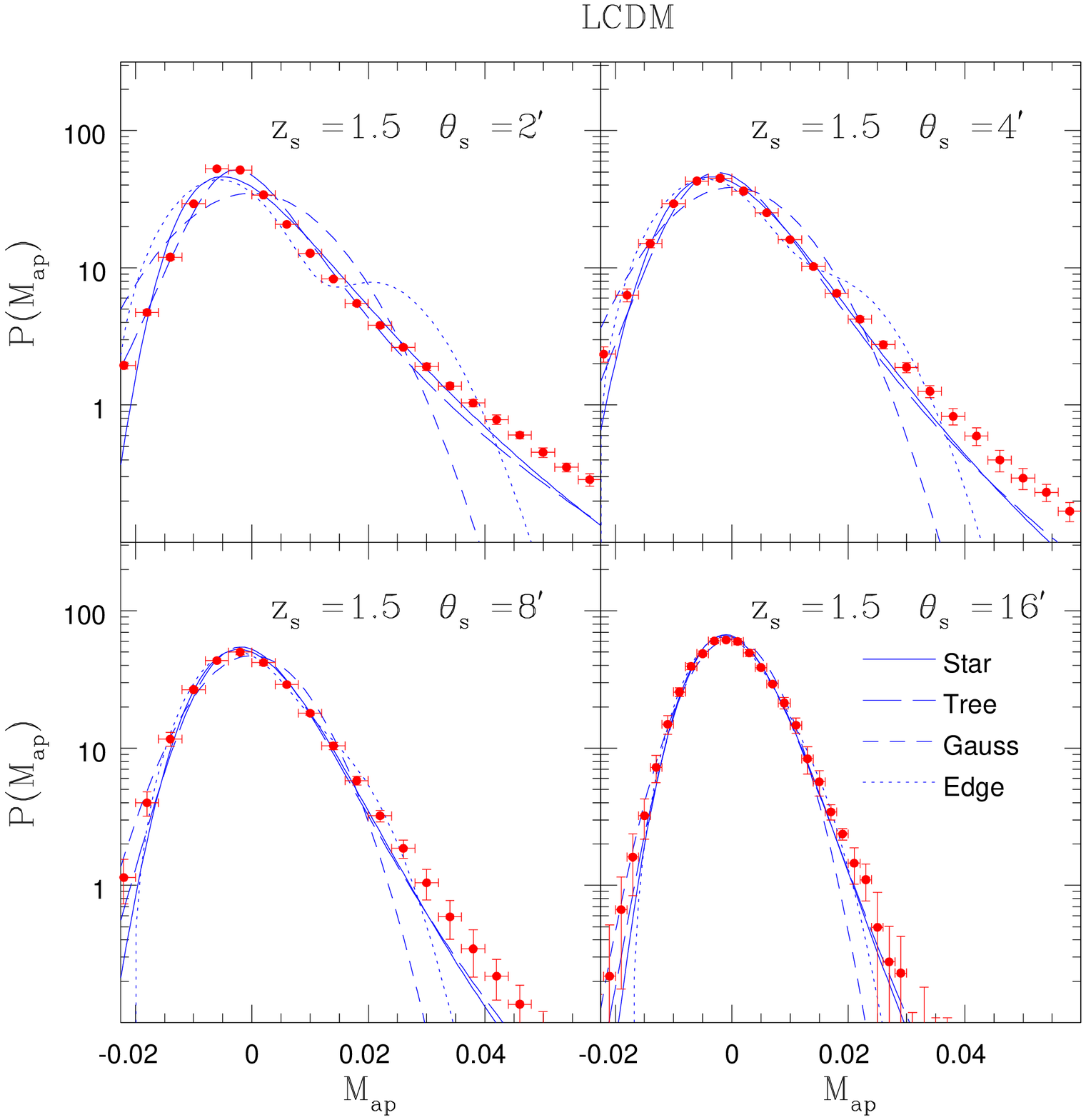}}
\caption{As in previous figure but for source redshift $z_s = 1.5$.}
\label{pdf_map_lcdmz1d5.eps}
\end{figure}

\begin{figure}
\protect\centerline{
\epsfysize = 3.25truein
\epsfbox[25 150 588 715]
{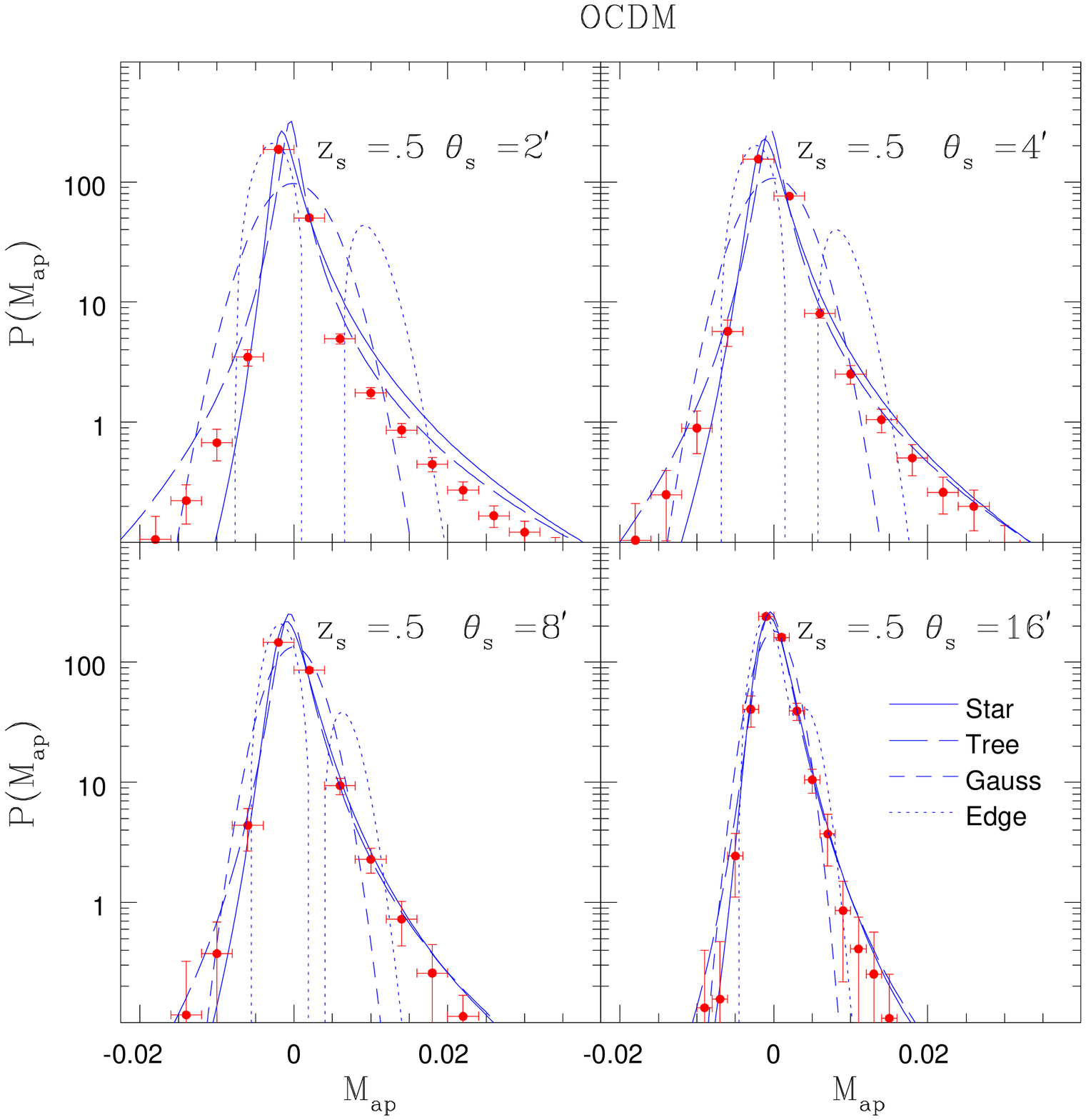}}
\caption{As in previous figure but for source redshift $z_s = 0.5$ and 
the OCDM cosmology.}
\label{pdf_map_ocdmd5.eps}
\end{figure}

\begin{figure}
\protect\centerline{
\epsfysize = 3.25truein
\epsfbox[25 150 588 715]
{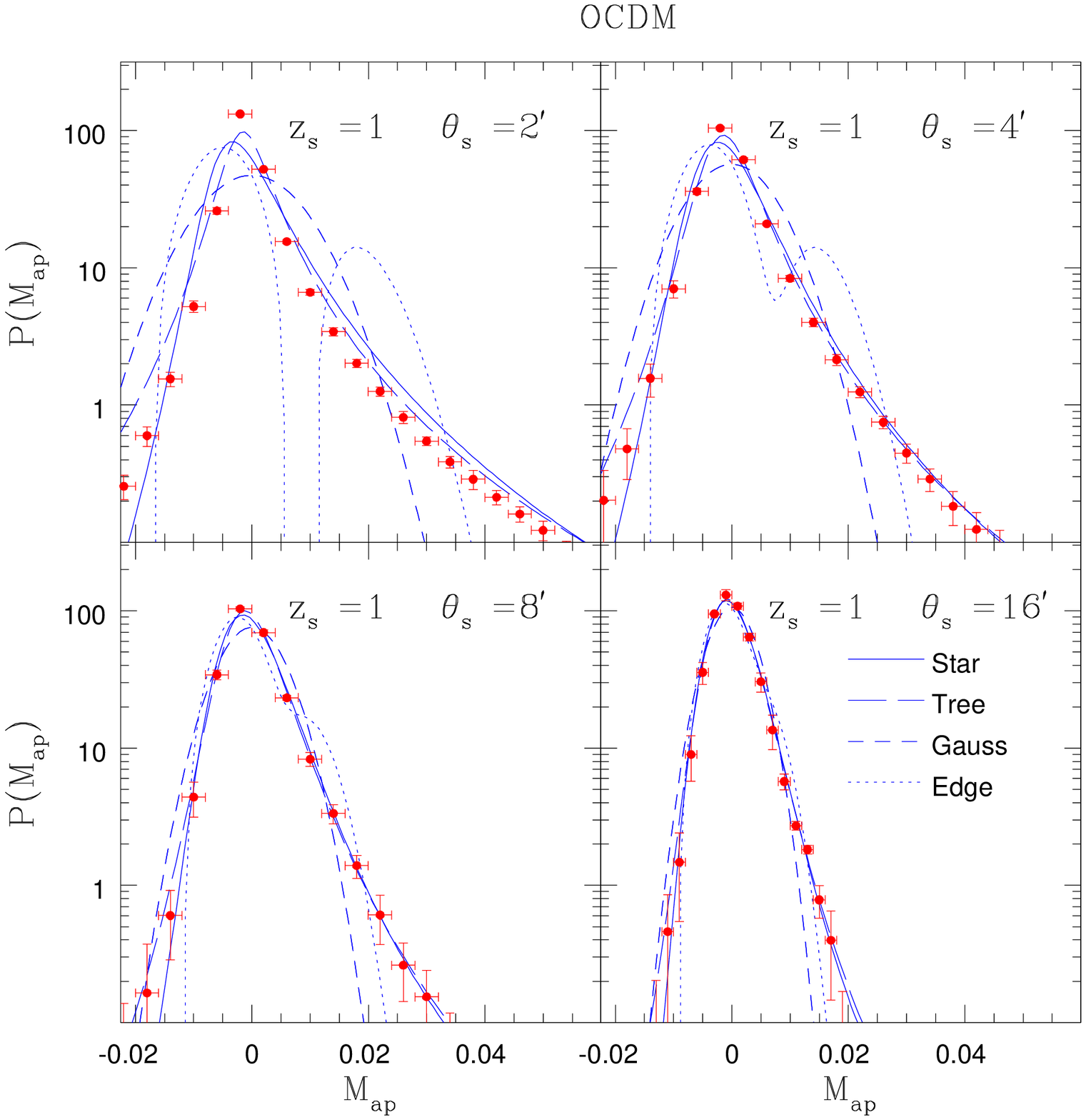}}
\caption{As in previous figure but for source redshift $z_s = 1$.}
\label{pdf_map_ocdm1d.eps}
\end{figure}

\begin{figure}
\protect\centerline{
\epsfysize = 3.25truein
\epsfbox[25 150 588 715]
{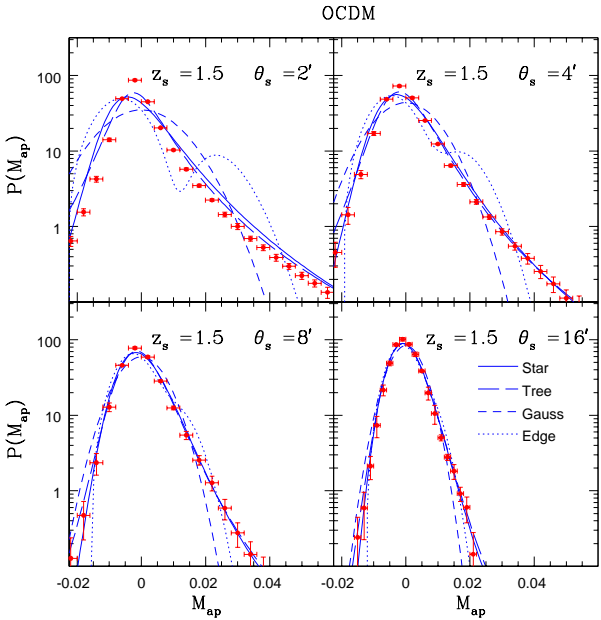}}
\caption{As in previous figure but for source redshift $z_s = 1.5$.}
\label{pdf_map_ocdmz1d5.eps}
\end{figure}

\subsection{Probability Distribution Function for Aperture-Mass}
\label{Probability Distribution Function for Aperture-Mass}

As discussed in Sect.~\ref{Skewness of Aperture-Mass}, the aperture-mass 
$\Map$ shows several advantages over the shear components $\gamma_{is}$.
In particular, it probes a narrower range of scales and it is not an even
quantity. Therefore, we plot in 
Figs.~\ref{pdf_map_lcdmzd5.eps} - \ref{pdf_map_ocdmz1d5.eps} the pdf 
$\cP(\Map)$ of the aperture-mass. We again consider both LCDM and OCDM
cosmologies and sources redshifts $z_s=0.5, 1$ and $1.5$. However, we now
study the smoothing angles $\theta_s=2',4',8'$ and $16'$. Indeed, as shown
in Fig.~\ref{FigD2k} the aperture-mass probes smaller scales than the shear
for a same angular radius, hence the limitation from the finite resolution of
numerical simulations shifts to larger angles. On the other hand, we now plot
the analytical predictions of both the minimal tree-model 
(\ref{phiom})-(\ref{tauom}) (long dashed lines) and the stellar model 
(\ref{phiXstellar}) (solid lines). Indeed, contrary to the convergence
(see Barber, Munshi \& Valageas 2003) or the shear, the predictions of both models
significantly differ at smaller angular scales for the large negative 
$\Map$ tail. As explained in Sect.~\ref{The tail}, this behaviour can be
understood on simple theoretical grounds. At very small angles 
($\theta_s \la 0.4'$) the pdf $\cP(\Map)$ predicted by the stellar model
shows a singular behaviour and it vanishes for $\Map < 0$. At the scales of
pratical interest displayed in Figs.~\ref{pdf_map_lcdmzd5.eps} - 
\ref{pdf_map_ocdmz1d5.eps} the pdf is still regular but we can see a hint of
the trend towards this singularity as the falloff at negative $\Map$ becomes
increasingly steep for smaller scales. By contrast, the pdf obtained from
the minimal tree-model which does not suffer from this singular behaviour
shows a smoother cutoff at negative $\Map$ (although it remains sharper than
for positive $\Map$, in agreement with numerical simulations). Note that this
behaviour shows that no realizable density field can exactly obey the 
stellar model (whatever the coefficients $S_p$), while this remains an open 
question for the minimal tree-model.

We can see in the figures that the prediction from the minimal tree-model
shows a good agreement with the numerical simulations over all scales and
redshifts of interest. On the other hand, as expected from the previous
discussion the stellar model fails to reproduce the tail at negative $\Map$,
except for large angles and redshifts. However, it yields a reasonable
prediction for the pdf for $\Map>0$ (albeit the minimal tree-model fares 
slightly better in this domain too). Therefore, we can conclude that 
the minimal 
tree-model is a much better tool to study the statistics of the aperture-mass.
Fortunately, although the numerical computation is still more difficult and
more computer time consuming than for the stellar model, the function 
$\tau({\vec \vartheta})=\tau(\vartheta)$ introduced in eq.(\ref{tauom}) 
is now only one-dimensional which makes the computation much easier than
for the shear (where $\tau({\vec \vartheta})$ was truly 2-dimensional),
since the filter $\UMap({\vec \vartheta})$ obeys a radial symmetry.

We also display the Gaussian (dashed lines) with the same variance. Of course,
at large redshifts and angles the pdf becomes closer to Gaussian but we can
see that even for $\theta_s \sim 8'$ and $z_s \sim 1.5$ the deviations from
Gaussianity are clear. Moreover, at smaller angles and redshifts the Gaussian
completely fails to reproduce the pdf obtained from numerical simulations.
It cannot follow the sharp peak at $\Map \simeq 0$, the extended tails
and the asymmetry of the pdf. Thus, we can see that the departure from 
Gaussianity is much more important for $\cP(\Map)$ than for the pdf 
$\cP(\gamma_{is})$ of the shear studied in 
Sect.~\ref{Probability Distribution Function for Shear}. Therefore, the pdf 
$\cP(\Map)$ of the aperture-mass should provide a much more efficient tool 
to measure the deviations from Gaussianity than the pdf $\cP(\gamma_{is})$ of
the shear. Note that non-Gaussian features, like low-order moments for
instance, exhibit a strong $\Om$-dependence, mainly due to the presence of 
the normalising factor in eq.(\ref{kappa}) (e.g., Bernardeau et al. 1997). 
Therefore, the aperture-mass is a convenient tool to measure such cosmological
parameters as well as the properties of the underlying density field.
On the other hand, we also plot the Edgeworth expansion (dotted
lines) up to the first non-Gaussian term (skewness). As was the case for the
shear, we can see that this asymptotic expansion is of very little use, since
it only fares well at large angles and redshifts where the Gaussian is
already a reasonable description (although it somewhat improves the shape of
the pdf near its maximum) while as soon as there is a significant deviation
from the Gaussian it yields spurious oscillations which make it useless (in 
some domains it even gives negative values for $\cP(\Map)$). 

Attempts in modeling the pdf $\cP(\Map)$ were initiated by Reblinsky et al. 
(1999), where a halo model of clustering was assumed in order to compute the 
positive tail of the pdf. Indeed, the high-$\Map$ tail is very efficient in 
probing and mapping out large concentrations of dark mass in weak lensing 
surveys. On the other hand, a complete prediction of the pdf for all values 
of $\Map$, which would be difficult to obtain from a halo model which cannot
faithfully describe low-density regions and filamentary structures, gives 
a valuable insight in probing both overdense and underdense regions in a 
statistical manner. We have shown that our method, which is based on the
many-body correlations rather than on a decomposition over halos, provides
such a model. In particular, the results presented above show that the
negative $\Map$-tail is very sensitive to the angular behaviour of the
correlation functions (in addition to their amplitude). Therefore, it provides
a direct probe of higher order correlations and it can help us
to discriminate among various models of gravitational clustering. In fact,
we have shown that it already rules out the stellar model even though this
model provides a very good description for the statistics of the convergence
(Barber, Munshi \& Valageas 2003) and the smoothed shear 
(Sect.~\ref{Probability Distribution Function for Shear}). 

Our results extend the work by Bernardeau \& Valageas (2000) who presented
a detailed study of the pdf $\cP(\Map)$ within both the quasi-linear regime
(where exact calculations are possible) and the highly non-linear regime 
(where they used the minimal tree-model). They also compared the predictions
of the minimal tree-model with numerical simulations for various cosmologies, 
with $z_s=1$ and $\theta_s=4'$. Here we have shown that the minimal tree-model
actually provides good predictions for $\cP(\Map)$ over all scales and 
redshifts of interest, from quasi-linear to highly non-linear scales.
Moreover, by introducing the stellar model we have shown that $\cP(\Map)$
could be used to probe the detailed angular behaviour of the many-body
correlations in addition to their amplitude. A more complete study of 
$\cP(\Map)$ with realistic source distributions as well as noise due to the
intrinsic ellipticity distribution will be presented elsewhere.


\begin{figure}
\protect\centerline{
\epsfysize = 1.75truein
\epsfbox[25 425 588 715]
{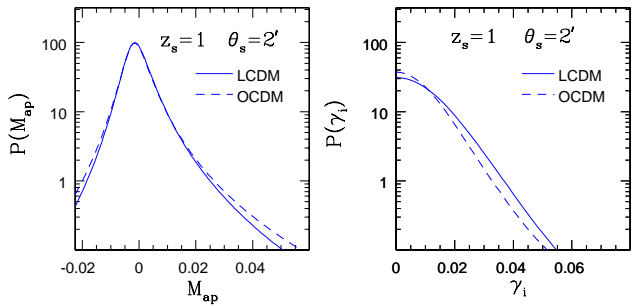}}
\caption{Probability distribution functions $\cP(\gamma_{is})$ of the smoothed 
shear component $\gamma_{is}$  (right panel) and $\cP(\Map)$ for the $\Map$
(left panel) are compared for two different cosmologies for a given smoothing 
angle $\theta_s =2'$ and source redshift $z_s = 1$. Solid lines correspond to 
LCDM cosmology and dashed lines correspond to OCDM cosmology.  }
\label{pdf_comp}
\end{figure}

\subsection{Comparison of the Probability Distribution Functions obtained
for both cosmologies}
\label{both cosmologies}

Finally, for the sake of completeness, we plot in Fig.~\ref{pdf_comp}
the pdfs obtained in both cosmological models for the shear and the 
aperture-mass. The difference between both models is rather small. Indeed,
these weak-lensing observables are mostly sensitive to the matter density
parameter $\Om$ and to the amplitude $\sigma_8$ of the density fluctuations
while the dependence on $\Ol$ is smaller. However, note that the comparison
is not so straightforward since both $\Ol$ and $\sigma_8$ differ between
the two models (and most of the change of the pdf comes from $\sigma_8$).
A more precise study of the dependence on cosmological parameters and on the
source redshift is left for future works.
%
%

\section{Discussion}
\label{Discussion}

Weak lensing surveys are being regularly used to probe the matter 
distribution and the underlying cosmology. However, the modeling of 
statistical quantities related to weak lensing surveys, such as convergence 
maps or related shear maps, is rather difficult because of the non-linear 
nature of the underlying density field they try to probe. Besides, the
non-linearity grows at smaller angular scales which should dominate the 
signal in forthcoming surveys. In a series of recent papers we have shown 
how to combine a widely used parameterization of the non-linear matter 
power-spectrum with various simple models of the matter density correlation 
hierarchy. This allows a detailed description of weak lensing statistics.

In the present paper we mainly focus on the statistical properties of the
weak lensing shear components $\gamma_{is}$ and the closely related 
aperture-mass $\Map$.
%
%
These two quantities can be computed from observational surveys in a direct 
manner from galaxy ellipticities (in the weak lensing regime). By contrast, 
extracting convergence maps is more difficult as it involves a non-local
inversion problem and exhibits a mass sheet degeneracy. However, we must note
that the measure of shear statistics still remains a difficult task because
of the non-trivial survey topology (a fraction of the survey area has to be 
excluded from the analysis because of image defects, bright stars,..).
Thus, rather than computing the shear or the aperture-mass realized over
many complete circles on the sky, one usually estimates low-order correlation 
functions of the shear which are next integrated in order to obtain low-order
moments of the smoothed shear or aperture-mass. However, we do not study
these points in this article.
%
%
We investigate two simple analytical descriptions of the matter density
field: the minimal tree model and the stellar model which we introduced in
Valageas, Barber \& Munshi (2004). Both models give identical results for the 
statistics of the 3-d density contrast smoothed over spherical cells and only 
differ by the detailed angular dependence of the many-body density 
correlations.

In a previous study (Barber, Munshi \& Valageas 2004) we have shown that both models also
give almost identical results for the smoothed convergence $\kappa_s$. In 
agreement with earlier works, this shows that the pdf $\cP(\kappa_s)$ 
(or its moments) provides a robust measure of the pdf $\cP(\delta_R)$. This 
also enables one to measure in a robust fashion the underlying cosmological 
parameters as well as the amplitude of the density correlations. In the 
present study we extend such calculations to more intricate compensated 
filters which can directly be constructed from shear maps: the smoothed shear 
components $\gamma_{is}$ and the aperture-mass $\Map$. As these 
observables involve more intricate filters we can expect the dependence on 
the angular behaviour of the many-body correlations to be larger than for
the smoothed convergence. We find that both models actually yield rather 
close results for the smoothed shear components and the positive tail 
of the aperture-mass. Moreover, we obtain a good agreement with numerical
simulations over all scales and redshifts of practical interest. Therefore, in 
this domain the shear and the aperture-mass provide again a robust constraint
on cosmological parameters and the statistics of the smoothed 3-d density
contrast $\delta_R$. Besides, we found that $\cP(\Map)$ shows a stronger
departure from the Gaussian than $\cP(\gamma_{is})$, so that the aperture-mass
appears to be a very useful tool in this respect.

On the other hand, we also note that at small angles ($\theta_s \la 2'$) the 
tail of the pdf $\cP(\Map)$ for negative $\Map$ shows a strong variation 
between both models. Whereas the minimal tree-model provides a good 
description of the numerical data down to the smallest scales available to us
the stellar model shows a significant discrepancy (while the part of the pdf 
over $\Map>0$ remains reasonable). The stellar model actually breaks down at 
$\theta_s \la 0.4'$ for $\Map<0$. This clearly indicates that the 
aperture-mass statistics can be used as a very precise probe of the high-order
correlation hierarchy. Contrary to the smoothed convergence $\kappa_s$, 
it is not only
sensitive to the amplitude of the many-body correlations but also to their
detailed angular behaviour. Thus, it provides a complimentary tool to
the smoothed convergence or shear components. From a theoretical point of 
view, we can note that this behaviour also means that no physical density 
field can be exactly described by the stellar model (while this remains an 
open issue for generic minimal tree-models) although it provides a very 
interesting and convenient approximation which works very well for both 
the smoothed convergence and the smoothed shear components. This shows
that the shape of the pdf $\cP(\Map)$ can actually rule out sensible
models of gravitational clustering.

Combined with our previous results for shear and convergence statistics
(Valageas, Barber \& Munshi 2004, Barber, Munshi \& Valageas 2004) we have 
developed a very powerful technique to analyse weak lensing survey results.
Although in our present calculations (both analytical and numerical)
we have not included observational details like the finite
width in source distribution or the noise due to the
intrinsic ellipticity distribution of galaxies, they can be incorporated
easily in our computations. A detailed and more elaborate comparison
will be presented elsewhere when such simulations become available.

Several interesting numerical issues became clear from our study.
For a given angular scale $\theta_s$ compensated filters like $\UMap$ pick up
more contributions from higher wavenumbers as compared with tophat filters.
This makes them more difficult to study numerically at smaller angular scales, 
as the finite resolution of the simulations starts to play a role. 
This is also the regime where $\Map$ statistics are very sensitive to
the detailed analytical modeling of the correlation hierarchy. 
Our work serves as a precurser to studies using simulations with much finer 
resolutions. However, using such compensated filters also means that finite 
volume effects are much less pronounced as compared with tophat filters 
(see Munshi \& Coles (2002) for a detailed analysis of various spurious
results in determination of $\Map$ statistics).

Most analytical studies in the non-linear regime use a halo model 
(for detailed predictions of a halo model see Takada \& Jain 2003).
In principles, halo models (see Cooray \& Sheth 200 for a review) can predict 
the higher-order correlation functions in the highly non-linear regime.
Indeed, on small scales the latter are set by the density profile of the halos
as the $p-$point correlations are dominated by the contribution associated
with all $p$ points being within the same halo. However, it is interesting 
to note that the neglect of substructures may lead to larger inaccuracies 
for high-order statistics. On the other hand, at intermediate scales one 
also probes the correlations among different halos which introduces new 
unknowns and makes explicit calculations cumbersome. This is important for 
handling projection effects which involve the mixing of various scales. 
Finally, such a model for the density field is not well-suited to describe 
the low-density and underdense regions (e.g., voids, filaments) which are 
outside virialized objects. Our method follows a completely different approach
based on the high-order correlation functions themselves rather than on a
decomposition of the matter distribution over a population of virialized halos.
Clearly such an approach provides an independent complimentary scheme
and will be useful for various numerical cross-checks.

The simulation technique that we employ is completely different from 
the more popular ray tracing methods. Our studies therefore not only
provide a good test for analytical results but it is also a good 
consistency test for such new simulation techniques.

\section*{acknowledgments}

DM acknowledges the support from PPARC of grant
RG28936. It is a pleasure for DM to acknowledge many fruitful
discussions with members of Cambridge Leverhulme Quantitative
Cosmology Group including Jerry Ostriker and Alexandre Refregier. 
This work has been supported by PPARC and the numerical work carried
out with facilities provided by the University of Sussex. AJB was
supported in part by the Leverhulme Trust. The original code for the
3-d shear computations was written by Hugh Couchman of McMaster
University.

\end{document}